\documentclass[10pt,sigconf,letterpaper,nonacm]{acmart}

\renewcommand\footnotetextcopyrightpermission[1]{}
\renewcommand\footnotetextcopyrightpermission[1]{} 
\def\BibTeX{{\rm B\kern-.05em{\sc i\kern-.025em b}\kern-.08emT\kern-.1667em\lower.7ex\hbox{E}\kern-.125emX}}
    
\usepackage{nicefrac}
\usepackage{siunitx}
\usepackage{array,framed}
\usepackage{booktabs}
\usepackage{
  color,
  float,
  epsfig,
  wrapfig,
  graphics,
  graphicx,
  subcaption
}

\usepackage{textcomp,amssymb}
\usepackage{setspace}
\usepackage{latexsym,fancyhdr}
\usepackage{enumerate}
\usepackage{graphics}
\usepackage{xparse} 
\usepackage{xspace}
\usepackage{multirow}
\usepackage{csvsimple}
\usepackage{balance}
\usepackage{amsmath}
\usepackage{color, colortbl}

\usepackage{
  tikz,
  pgfplots,
  pgfplotstable
}
\usepackage{hyperref}

\usetikzlibrary{
  shapes.geometric,
  arrows,
  external,
  pgfplots.groupplots,
  matrix
}
\usepackage{pifont}

\pgfplotsset{compat=1.9}

\usepackage{mathtools,}

\DeclareMathAlphabet{\mathcal}{OMS}{cmsy}{m}{n}

\DeclareGraphicsExtensions{%
    .png,.PNG,%
    .pdf,.PDF,%
    .jpg,.mps,.jpeg,.jbig2,.jb2,.JPG,.JPEG,.JBIG2,.JB2}
\usepackage{xparse}
\newcommand{\bnm}{\begin{newmath}}
\newcommand{\enm}{\end{newmath}}

\newcommand{\bea}{\begin{eqnarray*}}%
\newcommand{\eea}{\end{eqnarray*}}%

\newcommand{\bne}{\begin{newequation}}
\newcommand{\ene}{\end{newequation}}

\newcommand{\bal}{\begin{newalign}}
\newcommand{\eal}{\end{newalign}}

\newenvironment{newalign}{\begin{align}%
\setlength{\abovedisplayskip}{4pt}%
\setlength{\belowdisplayskip}{4pt}%
\setlength{\abovedisplayshortskip}{6pt}%
\setlength{\belowdisplayshortskip}{6pt} }{\end{align}}

\newenvironment{newmath}{\begin{displaymath}%
\setlength{\abovedisplayskip}{4pt}%
\setlength{\belowdisplayskip}{4pt}%
\setlength{\abovedisplayshortskip}{6pt}%
\setlength{\belowdisplayshortskip}{6pt} }{\end{displaymath}}

\newenvironment{newequation}{\begin{equation}%
\setlength{\abovedisplayskip}{4pt}%
\setlength{\belowdisplayskip}{4pt}%
\setlength{\abovedisplayshortskip}{6pt}%
\setlength{\belowdisplayshortskip}{6pt} }{\end{equation}}

\newcounter{ctr}

\newcounter{mytable}
\def\mytable{\begin{centering}\refstepcounter{mytable}}
\def\endmytable{\end{centering}}

\newcounter{myfig}
\def\myfig{\begin{centering}\refstepcounter{myfig}}
\def\endmyfig{\end{centering}}

\newlength{\saveparindent}
\newlength{\saveparskip}
\newcommand{\E}{{\rm I\kern-.3em E}}

\renewcommand{\eqref}[1]{\mbox{Equation~(\ref{#1})}}

\def \part {part}

\renewcommand{\paragraph}[1]{\vspace*{6pt}\noindent\textbf{#1}\;}

\def \blackslug{\hbox{\hskip 1pt \vrule width 4pt height 8pt
    depth 1.5pt \hskip 1pt}}
\def \qed{\quad\blackslug\lower 8.5pt\null\par}

\newcounter{mynote}[section]

\newcommand\ignore[1]{}

\newcounter{rcnote}[section]

\newcounter{mrnote}[section]

\newcounter{fknote}[section]

\newcounter{anote}[section]

\DeclareMathSymbol{\mlq}{\mathord}{operators}{``}
\DeclareMathSymbol{\mrq}{\mathord}{operators}{`'}

\newcommand{\rhf}[2]{R_{f, \gamma}}

\DeclareDocumentCommand{\edist}{o o}{
  \ensuremath{
    \IfNoValueTF{#1}{{d}}{{\sf d}(#1,#2)}
  }
}

\newcommand{\olrk}[1]{\ifx\nursymbol#1\else\!\!\mskip4.5mu plus 0.5mu\left(\mskip0.5mu plus0.5mu #1\mskip1.5mu plus0.5mu \right)\fi}

\NewDocumentCommand{\indseq}{ O{1} O{r} }{{#1}\ldots {#2}}

\newcommand{\mypara}[1]{\vspace{0.07cm}\noindent{\bf {#1}:}~}

\newcommand{\eg}{{\it e.g.,}\xspace}
\newcommand{\ie}{{\it i.e.,}\xspace}

\definecolor{Gray}{gray}{0.9}

\newcommand{\name}{{VidHoc}\xspace}

\newcommand{\xu}[1]{{\color{black}{#1}}}

\newcommand{\fillme}{{\bf XXX}\xspace}

\newcounter{packednmbr}
\newenvironment{packedenumerate}{\begin{list}{\thepackednmbr.}{\usecounter{packednmbr}
\setlength{\itemsep}{0.5pt}\addtolength{\labelwidth}{-4pt}\setlength{\leftmargin}{2.5ex}\setlength{\listparindent}{\parindent}\setlength{\parsep}{1pt}\setlength{\topsep}{2pt}}}{\end{list}}
\newenvironment{packeditemize}{\begin{list}{$\bullet$}{
\setlength{\itemsep}{0.5pt}\addtolength{\labelwidth}{-4pt}\setlength{\leftmargin}{2.5ex}\setlength{\listparindent}{\parindent}\setlength{\parsep}{1pt}\setlength{\topsep}{2pt}}}{\end{list}}

\newcommand{\tightsection}[1]{\vspace{-0.05cm}\section{#1}\vspace{-0.05cm}}
\newcommand{\tightsubsection}[1]{\vspace{-0.05cm}\subsection{#1}\vspace{-0.05cm}}

\begin{document}
\fancyhead{}
\def\thetitle{\scalebox{0.98}{\huge Enabling Personalized Video Quality Optimization with \name}}
\title{\thetitle}


\def\theauthor{\scalebox{0.98}{\large Xu Zhang, Paul Schmitt*, Marshini Chetty, Nick Feamster, Junchen Jiang}}

\def\theinstitution{\scalebox{0.98}{\large University of Chicago, University of Hawai`i at M\=anoa*}}

\author{\theauthor}
\affiliation{%
  \institution{\theinstitution}
  \country{}
}

\date{}


\begin{abstract}
Conventional approaches to optimize user quality of experience (QoE) for
    online video streams typically consider certain aspects of heterogeneity,
    including the content of the video and the device being used to watch the
    video.  Yet, existing video delivery systems have ignored one of the biggest
    sources of heterogeneity in video QoE: the quality preferences of
    \emph{individual} users.  We demonstrate, using three independent
    datasets, that users exhibit considerable heterogeneity in QoE, even when
    watching the same video, on the same device, with the same video quality.
    Previous work has suggested that such heterogeneity can be exploited to
    improve QoE, and yet the challenge of automatically learning user QoE
    preferences without imposing undue burden on the user has remained an open
    problem.  To address this problem, this paper presents \name, a
    bandwidth-control system which, for each new user, automatically adjusts
    throughput during each video session to (1)~learn the user's unique QoE
    model and (2)~optimize resource allocation in accordance with these
    preferences.  One significant challenge involves quickly reducing model
    uncertainty (which requires subjecting the user to sub-optimal video
    quality) without unacceptably degrading the user's overall experience.
    To address this challenge, \name frames this problem in terms of
     the well-studied framework of regret minimization: for each video
    session, \name selects video quality that minimizes the
    ``regret'' on the current session's QoE (\ie gap to the best possible
    quality) and on future video sessions' QoE (\ie inaccuracy of QoE model).
    To evaluate \name in a real-world setting, we implemented a browser-based
    user-study platform, deployed it to 15 users, and tested \name against
    baseline resource allocation strategies
    for four months.  The results show that \name can
    improve QoE (measured in engagement) by 13.9\% with the same bandwidth or
    save 17.3\% bandwidth while maintaining the same QoE.
\end{abstract}

\maketitle


\tightsection{Introduction}
\label{sec:intro}

Video streaming systems strive to maximize user-perceived quality of experience (QoE) within the constraints imposed by limited network resources.
Prior work has focused on optimizing  bandwidth allocation at bottlenecks or bitrate adaptation algorithms (ABR) on the video player.
These techniques are based on {\em QoE models} that are built {\em offline} to characterize relationships between QoE and common video quality metrics (such as video bitrate, rebuffering time, bitrate switch probability).
Given these relationships, existing systems seek to optimize QoE by improving the underlying quality metrics.
Traditionally, the QoE models leverage general features, such as video playback modalities (\eg smartphone, desktop, TV), video types (\eg live, VoD), video content (\eg sports, news).
For each combination of these generic features, a QoE model is developed offline (via user study~\cite{zhang2021sensei, duanmu2019knowledge} or analysis of historical data~\cite{gao2020personalized}) and then applied to new users.

However, these traditional QoE models fail to capture the differences among {\em individual} users. 
Indeed, the literature has shown that in terms of how quality affects QoE, there is substantial heterogeneity across users.
Our analysis over three datasets~\cite{duanmu2019sqoe4, zhang2021sensei} shows that a one-second rebuffering event can make the QoE rating drop from 2\% to 51\% across the users (\S\ref{sec:moti:hetero}). 
Moreover, different users are more sensitive to different aspects of video quality.

To exploit this user heterogeneity, previous efforts have tackled two questions.
(1) How to efficiently build QoE models using {\em offline} user studies and data analysis. 
A common approach is to let recruited participants first watch videos (often identical content) with manually selected quality issues and then collects user feedback to learn the per-user QoE models~\cite{zhang2021sensei,duanmu2016sqi,duanmu2019knowledge}. 
(2) How to leverage {\em known} heterogeneity among different users' video quality preferences  to optimize QoE on an individual basis.
For instance, client-side ABR logic~\cite{mao2017neural, yan2020learning, yin2015control} can be customized based on per-user QoE models, and bandwidth can be better allocated across users with heterogeneous QoE models~\cite{nathan2019end}.

However, prior work does not address {\em how to automatically learn a QoE model for each new user while simultaneously optimizing the user's QoE?}
For a content provider or Internet Service Provider (ISP), it is impractical to ask users to repeatedly watch identical videos at various quality levels, as in previous user studies.
Moreover, these user studies often show users poor video quality, which helps to elicit QoE preferences but is directly at odds with service providers' goal to maximize QoE in each video session.
Thus, the potential benefits of leveraging cross-user QoE heterogeneity remain unrealized in practice. 

In this paper, we explor a new approach: when a new user joins a video service, we customize the quality of the videos that the user elects to watch, and automatically gather user QoE feedback (\eg user engagement) over time to gradually learn a more accurate per-user QoE model, while applying the latest QoE model to optimize the user's QoE.
Our approach effectively builds a QoE model for each new user while they watch videos in their natural environment.

This approach spares the users from watching curated videos repeatedly, but we must nonetheless carefully select the quality of a user's video sessions, such that we minimize the potential negative impact on user-perceived QoE as we try to learn users' QoE preferences.
There are two seemingly natural solutions that might lower the user annoyance.
The first solution is to passively measure user QoE under their natural video quality, \ie show the videos without deliberately adding quality issues.
This method does have minimum impact on QoE in each session, but it might take a long time to accumulate enough samples to learn an accurate per-user QoE model. 
The second solution is to select the quality for a small number of videos such that their QoE feedback will maximally increase the accuracy of the per-user QoE model after a small number of sessions.
This approach does reduce the number of QoE samples, but users might experience low QoE during the process; 
after all, the QoE samples that are more informative to QoE modeling tend to have lower
quality.
To facilitate a large-scale deployment of this approach, we use {\em user engagement time} as the metric of QoE instead of soliciting the user rating (\eg push a button to give a five-start rating) by an exit survey after a video session.
User engagement time (\ie the ratio of the video viewing time over the video length) is currently widely used in industry, mostly because the revenue highly depends on it. 
Compared with the user rating, collecting user engagement time does not need the efforts from users, which makes it much easier to automate the process.

We address this challenge in the concrete context of a new system, called {\em \name}, which controls bandwidth, and thus flow throughput, during a video session.
Since video player ABR algorithms are influenced by throughput history, 
\name controls video quality by dynamically adjusting available bandwidth while maintaining a fixed average bandwidth during a video session.
We can leverage bandwidth tuning to either optimize user QoE or to solicit QoE feedback to better understand the user's QoE preferences. 
After the users watch videos, we measure QoE by the engagement time~\cite{balachandran2012quest, dobrian2011understanding}.
The measurement can be at the side of an ISP or or Content Delivery Network (CDN) that controls bandwidth of the bottleneck link and monitor the duration of the video streaming flow.

\name balances the training videos' QoE and user feedback to contribute to its individualized QoE model training by framing the problem in terms of the well-studied {\em regret-minimization} framework.
In each video session, we seek to minimize the ``regret'' of the video shown to the user---intuitively, the ``regret'' captures both the gap to optimal quality on the current session's QoE and the gap to optimal quality for future video sessions' QoE due to inaccuracy of QoE model.
A natural benefit of this formulation is that it does not need to pre-determine a set of curated videos to profile a user's QoE model; 
instead, we dynamically determine a quality based on both the uncertainty of the current QoE model as well as the possible video quality in the current video session.
Over time, as more QoE samples are collected, the per-user QoE model becomes more accurate, naturally diminishing the need for QoE modeling.

A practical challenge facing \name is that, due to the proprietary nature of the video player, changing bandwidth at the bottleneck link may induce different player actions and user-perceived quality. 
For instance, if \name decides to measure a user's QoE when the resolution drops from 1080p to 360p, it will lower the available bandwidth, but as a result, player may either drop the resolution from 1080p to 360p, or continue playing at 1080p but experience a rebuffering event instead.
To address this issue, we change our goal from determining a specific video quality pattern that minimizes regret, to dynamically creating a bandwidth schedule that can minimize the \emph{expected} regret (\eg gap between optimal QoE and QoE model uncertainty) over a distribution of possible video quality resulting from the chosen bandwidth schedule.

To test \name in a real-world setting, we built and deployed a user-study platform to fifteen crowdsourced participants over four months.
Each participant installed a browser extension that monitors the YouTube player status (\eg video quality, user engagement) and dynamically controls a user's video downloading rate to induce specific video quality perceived by users.
Using this platform, we ran randomized A/B tests to compare \name with various baselines in the real-world settings.
Our results show that \name can yield 9.2--21.6\% higher average user engagement over various baselines (including today's default system) during a user's first 60 sessions.

This paper presents two main contributions:
\begin{packedenumerate}
    \item By formulating the problem as regret minimization, we demonstrate that it is possible to automatically learn a new user's unique QoE models in an online fashion, with minimal impact on the user's QoE, while quickly providing the ability to optimize QoE for the user. 
    \item To test the solution in practice, we have developed a bandwidth-control system and a new user-study platform that automates the testing of various QoE profiling algorithms on recruited participants. 
    In a study of 15 users over four months, we show that we can improve the QoE by 13.9\% while maintaining the same bandwidth usage, or maintain the same QoE while using 17.3\% less bandwidth. 
\end{packedenumerate}


\tightsection{Motivation}
\label{sec:moti}


First, we motivate our work with an analysis of user heterogeneity over three video QoE datasets. 
We show that in terms of how various video quality metrics affect user QoE, there is significant user heterogeneity.
Unlike prior work~\cite{chang2018active, zhao2019qoe} that studies user QoE in one dataset, our analysis presents a more panoramic view across two QoE metrics (user rating and user engagement) in different settings (in-lab and crowdsourced surveys with curated test videos and open-world tests without video duration).
Our findings informed the design requirements of our system, \name.

\tightsubsection{Datasets and method}

We investigate user heterogeneity in a corpus of three datasets (Table~\ref{tab:dataset}).
Two of them are opensource datasets: Waterloo~\cite{duanmu2019knowledge} and Sensei~\cite{zhang2021sensei}.
Both datasets were collected in controlled environments where the recruited participants were asked to watch the same set of original videos, where each played multiple times but at different quality (in terms of average bitrate, bitrate switches, and rebuffering stalls), and then rated their experience (QoE) after each video in the video session exit surveys.
The datasets record the quality metrics of each played video and the corresponding user rating.

To complement these datasets, we also collected our own dataset from 15 participants recruited from Amazon Mechanical Turk~\cite{mturklink} and Prolific~\cite{prolificlink}.
The participants install a browser plugin that passively monitors the YouTube player status to accurately measure the video quality of each YouTube video session. 
This third dataset is different in two aspects.
{\em (i)} Instead of watching curated videos in a controlled setting, participants can watch any videos at any time they wish.
{\em (ii)} Instead of user ratings, QoE is measured by user engagement (\ie the ratio of the video viewing time over the video length), which is frequently used in previous video QoE studies~\cite{balachandran2012quest, dobrian2011understanding}.
\S\ref{sec:impl} will describe the complete details of the collection of the third dataset. 

In the first two datasets (Waterloo and Sensei), the users 
were required to watch the same set of video content with the same viewing devices to eliminate the influence of video content.
In our own dataset, the content of a video and its assigned quality are independent, allowing us to mitigate other QoE-affecting factors by averaging over multiple video sessions.

\begin{table}[t]
\vspace{1em}
\small
\begin{tabular}{|l|l|l|l|}
\hline
Dataset              & \# sessions & \# users & QoE metric                                                 \\ \hline
Waterloo~\cite{duanmu2019knowledge}      &  1,350              & 31       & Exit survey rating                                           \\ \hline
Sensei~\cite{zhang2021sensei}               & 1,600              & 15       & Exit survey rating                                               \\ \hline
\name & 3,750              & 15       & \begin{tabular}[c]{@{}l@{}}User engagement\\\end{tabular} \\ \hline
\end{tabular}
\vspace{0.5cm}
\caption{Three datasets used to analyze user heterogeneity.
}
\label{tab:dataset}
\end{table}

We use a simple method to measure QoE user heterogeneity in these datasets. 
For each user, we measure the average QoE drop caused by a single quality issue: with vs. without a short rebuffering, low vs. high birate, and with vs without bitrate switches.
For instance, 
we consider a rebuffering event of 0.3-0.5 seconds (in Waterloo and \name datasets) or 1.0 second (in Sensei videos), and denote it as $Buf$.
To measure its effect on each user $i$,
we create two groups of QoE data: one group includes data of videos without
rebuffering, and the other group includes data with the event $Buf$, while the videos in both groups have no other quality issues (highest bitrate and no bitrate switch).

Then, we calculate the QoE difference $\Delta Q_{Buf}(i)$ caused by quality event $Buf$ on user $i$ by the difference in average QoE between the two group:
\begin{align}
   \Delta Q_{Buf}(i) = \text{ }&AvgQoE(\text{user } i\text{'s videos} \text{ with }Buf) - \nonumber \\
   &AvgQoE(\text{user } i\text{'s videos} \text{ without }Buf)
    \label{eq:qoedrop}
\end{align}
In the three datasets, video quality is chosen independently to the video content, so if the two groups contain enough data points when calculating  $\Delta Q_{Buf}(i)$, then we can minimize the impact of confounding effect of video content on QoE.

Similarly, we define $\Delta Q_{lowbitrate}(i)$ and $\Delta Q_{bitrateswitch}(i)$ to be the impact of low bitrate of $200-500 Kbps$
(the lowest bitrate level) and a bitrate switch of magnitude $1-1.5 Mbps$ on QoE of user $i$, respectively.
By restricting our analysis to a single aspect of video quality, these metrics allow us to compare users along each aspect of video quality separately. 
(To capture the complex interactions between user QoE and multiple aspects of video quality, we will consider more thorough QoE models in next sections.)

\begin{figure}[t]
      \begin{subfigure}[t]{1.0\linewidth}
      \captionsetup{justification=centering}
         \includegraphics[width=1.0\linewidth]{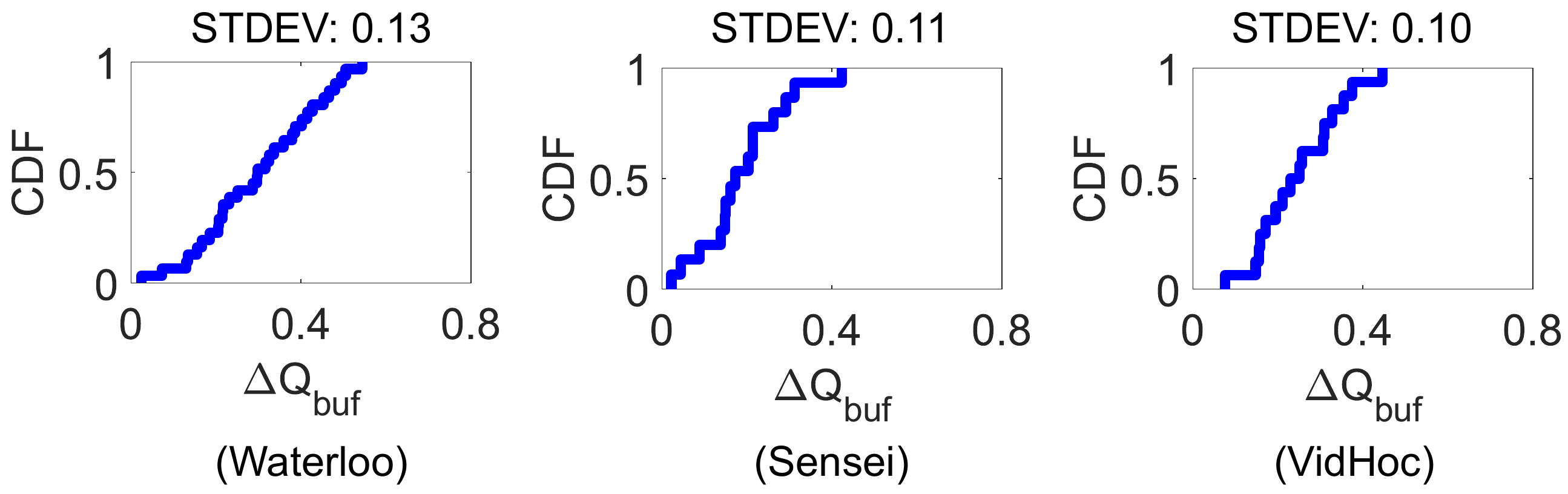}
         \vspace{-1.7em}
        \caption{Sensitivity to rebuffering}
        \label{subfig:rebuf}
      \end{subfigure}
          \vfill
          \vspace{0.8em}
          \begin{subfigure}[t]{1.0\linewidth}
      \captionsetup{justification=centering}
         \includegraphics[width=1.0\linewidth]{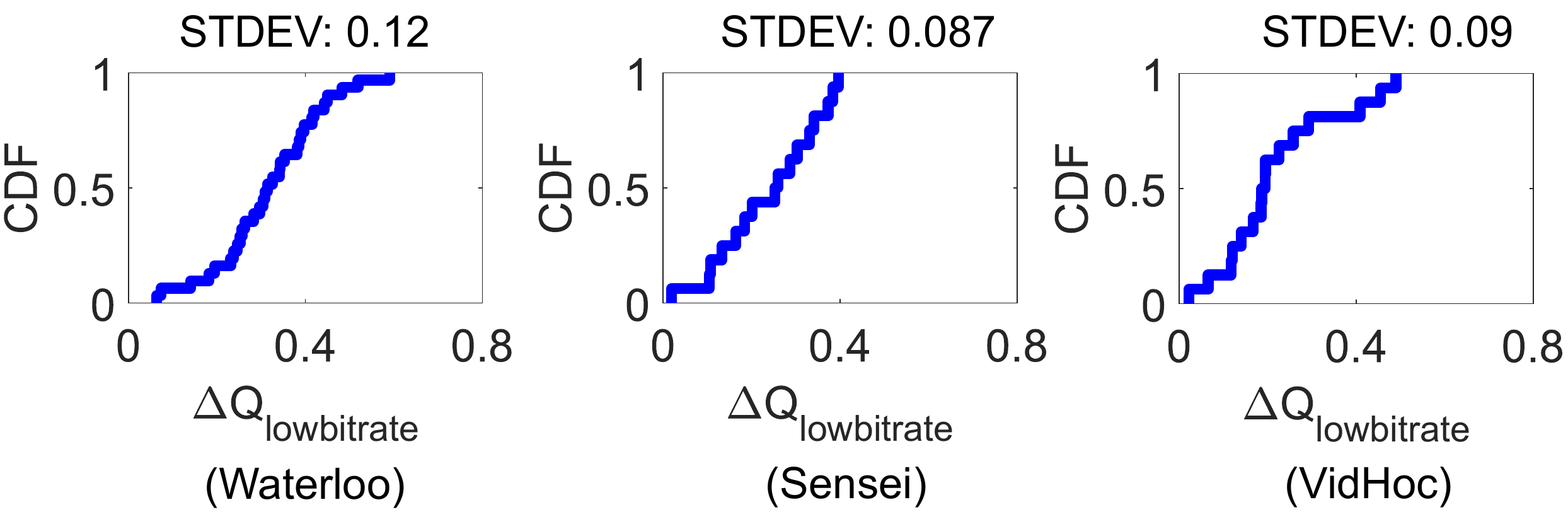}
         \vspace{-1.7em}
        \caption{Sensitivity to low bitrate}
        \label{subfig:rebuf}
      \end{subfigure}
          \vfill
          \vspace{0.8em}
          \begin{subfigure}[t]{1.0\linewidth}
      \captionsetup{justification=centering}
         \includegraphics[width=1.0\linewidth]{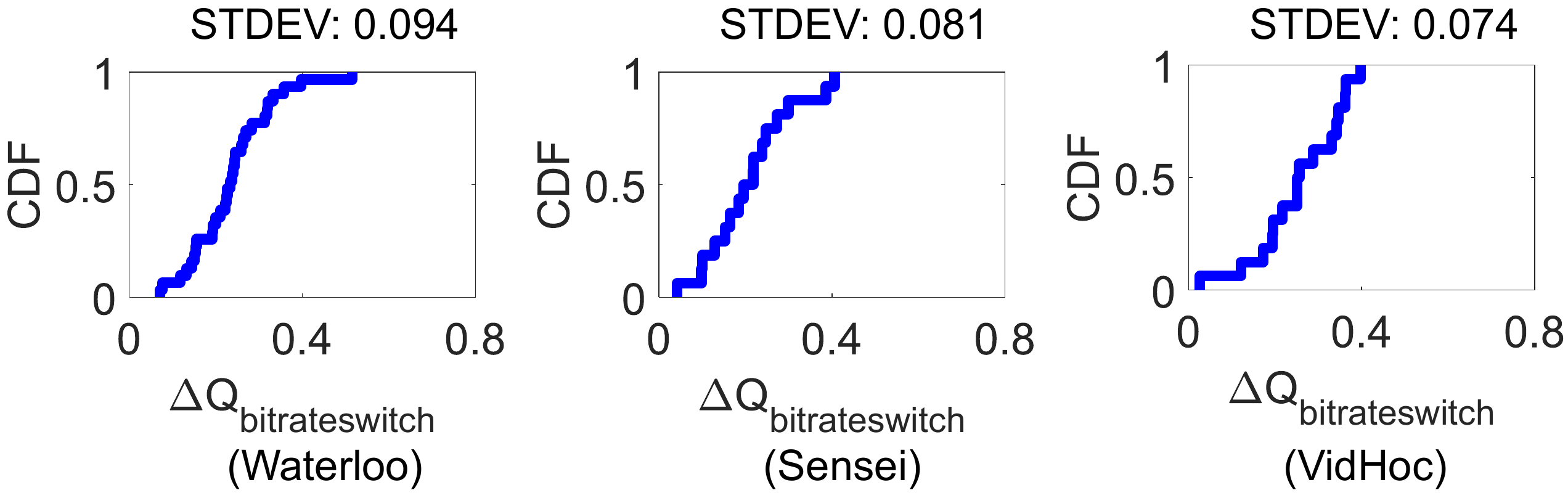}
         \vspace{-1.7em}
        \caption{Sensitivity to bitrate switch}
        \label{subfig:rebuf}
      \end{subfigure}
    \vfill
   \vspace{1em}
    \caption{There is a significant heterogeneity across users in terms of their sensitivity to a fixed low quality of 
    (a) a short rebuffering event of 0.3-0.5 seconds (in Waterloo and \name datasets) or 1.0 second (in Sensei videos), 
    (b) a fixed low average bitrate (200--500~Kbps lower than highest bitrate), and 
    (c) a bitrate switch (with a magnitude of 1--1.5~Mbps). 
    CDFs of QoE rating differences of different users in response to each quality issue in the Waterloo, Sensei, and \name datase.
}
\vspace{-0.2em}
    \label{fig:diff_react}
\end{figure}

\tightsubsection{Measuring user heterogeneity}
\label{sec:moti:hetero}

Figure~\ref{fig:diff_react} shows the distribution of $\Delta Q_{Buf}$, $\Delta Q_{lowbitrate}$, and $\Delta Q_{bitrateswitch}$ over the users in each dataset.
In terms of how sensitive QoE is to these three aspects of video quality, user heterogeneity is consistently significant.
We see that top 33\% users on average are 67\% more sensitive to the same quality difference than the the bottom 33\% of users.
We also observe similar trends of user heterogeneity between the two subjective QoE metrics (user rating vs. user engagement).

\begin{figure}[t]
    \includegraphics[width=1.0\linewidth]{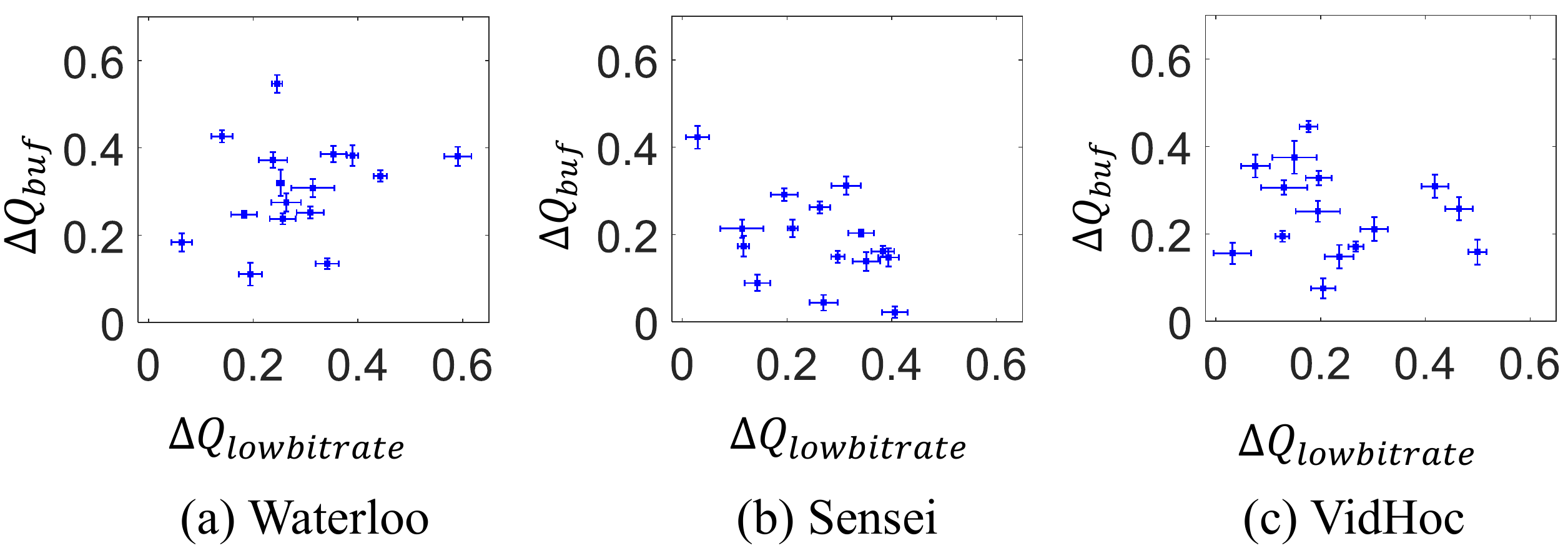}
    \vspace{-1.5em}
    \caption{Some users are more sensitive to rebuffering events, but some users are more sensitive to low bitrate.
    }
    \label{fig:types}
\end{figure}

\begin{figure}[t]
    \includegraphics[width=1.0\linewidth]{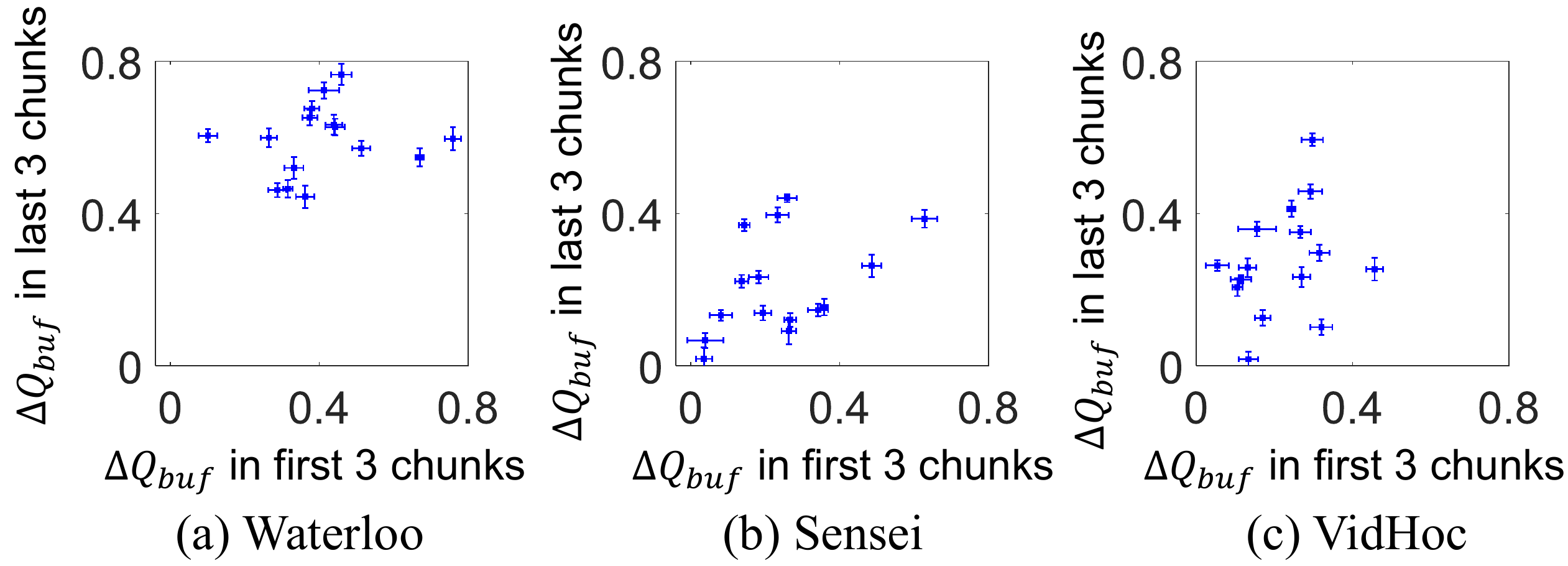}
    \vspace{-1.5em}
    \caption{Users' sensitivity to quality issues is time-dependent.
    Each dot is the mean and one standard deviation across 
    }
    \label{fig:memory}
\end{figure}


While Figure~\ref{fig:diff_react} measures user heterogeneity along one single quality aspect, Figure~\ref{fig:types} shows user heterogeneity along two quality aspects (rebuffering $\Delta Q_{Buf}(i)$ and low bitrate $\Delta Q_{lowbitrate}$).
For example, for 63.2\% of user pairs in Waterloo one user is more sensitive on one quality aspect but less sensitive on the other aspect. 
This suggests that the user heterogeneity observed in Figure~\ref{fig:diff_react} is not simply because some users are generally more sensitive than others; instead, users have different emphasis on various quality metrics.



Next, we show that one user might be more sensitive to video quality degradation depending on {\em when} in the video the degradation occurs. 
We select the video sessions with a rebuffering event of 0.3--1.0 seconds and divide them into two groups---videos where the rebuffering event occurs in the first 3 chunks and others where the rebuffering event occurs in the last 3 chunks. 
Figure~\ref{fig:memory} shows the QoE drop due to a rebuffering event in the first 3 chunks vs. the QoE drop due to a rebuffering event in the last 3 chunks.
In making this figure, we keep the average bitrate and the number of bitrate switches in a very small range (the difference is less than 12.4\%), so the QoE drops are primarily due to the rebuffering events and their positions in a video.
Across three datasets, we see that some users are more sensitive to ``early'' rebuffering and others are more sensitive to ``later'' rebuffering. (We observe similar effects with respect to low bitrate and bitrate switches too.)
A possible explanation is the memory of an event (\eg quality degradation) will fade over time, but how quickly the memory of that event fades differs across users.
This observation also suggests that when we build per-user QoE models, we should weight different parts in the video differently (\S\ref{sec:design}).

\begin{table}[t]
\footnotesize
\begin{tabular}{lrrr}
\toprule
Quality incident & \begin{tabular}[c]{@{}c@{}}{\bf Cross-{\em user}}\\ {\bf Heterogeneity}\end{tabular} & \begin{tabular}[c]{@{}c@{}}Cross-{\em device}\\ Heterogeneity\end{tabular} & \begin{tabular}[c]{@{}c@{}}Cross-{\em video}\\ Heterogeneity\end{tabular} \\ \midrule
\rowcolor{Gray}  Rebuffering & {\bf 0.33$\pm$ 0.03} & 0.41$\pm$ 0.03  & 0.34$\pm$ 0.01 \\ 
Avg bitrate & {\bf 0.37$\pm$ 0.02} & 0.55$\pm$ 0.06 & 0.31$\pm$ 0.03 \\ 
\rowcolor{Gray} Bitrate switches & {\bf 0.39$\pm$ 0.02} &  0.60$\pm$ 0.01 & 0.07$\pm$ 0.01 \\ \bottomrule
\end{tabular}
\vspace{1.2em}
\caption{The magnitude of user-level heterogeneity (standard deviation divided by the mean) is at least on par with the cross-device heterogeneity and cross-video heterogeneity. 
The confidence intervals are calculated by bootstrapping with 10\% random sampling (with replacement) over all QoE ratings.
}
\label{tab:on-par}
\vspace{-2.8em}
\end{table}

Finally, we put the cross-user heterogeneity in the perspective of other factors that might influence QoE.
The conventional belief is that the video content and the viewing device play an important role in a user's QoE preference.
Here, we calculate the cross-device and cross-video QoE heterogeneity in the Waterloo dataset as follows.
We first calculate the average QoE drops (Eq.~(\ref{eq:qoedrop})) due to the same short rebuffering event used in Figure~\ref{fig:diff_react} in each video or each device (HDTV, UHDTV, smartphone), and calculate the variance of these QoE drops across different videos or different devices.
Table~\ref{tab:on-par} shows the standard deviations (normalized by mean) across individual users, across videos, and across devices.
We see that the magnitude of user-level heterogeneity is at least on par with the device-induced difference and content-induced difference.



\mypara{Summary} The key takeaways from our analysis are:

\begin{packeditemize}
    \item Across three datasets that span different QoE metrics and survey methodologies, there is always significant user heterogeneity, in terms of how each quality aspect (rebuffering, low bitrate, and bitrate switch) affect user QoE.
    \item User heterogeneity is not as simple as some users being more sensitive to video quality than others; rather, users differ in terms of the aspect of video quality their QoE is most sensitive to and when in a video low quality occurs.
    \item The magnitude of heterogeneity in video quality preferences across individual users is on par with other well-known factors such as video content and viewing devices.
\end{packeditemize}

\tightsection{Online QoE modeling and optimization}
\label{sec:framework}


We have seen that there is substantial user heterogeneity in QoE.
Now, several previous efforts have proposed QoE optimization (ABR logic~\cite{huo2020meta} and bandwidth allocation~\cite{chen2015qoe}) that take advantage of {\em known} user preferences.
However, they are unable to optimize a new user right after the user joins, as one must first model the user's unique QoE preference.

To bridge the gap, in this section, we first formulate the general problem of {\em online QoE modeling and optimization} and elaborate why existing techniques designed for different-yet-related purposes are not well-suited for this framework. 

\tightsubsection{Framework}

\begin{figure}[t]
    \centering
    \includegraphics[width=0.9\linewidth]{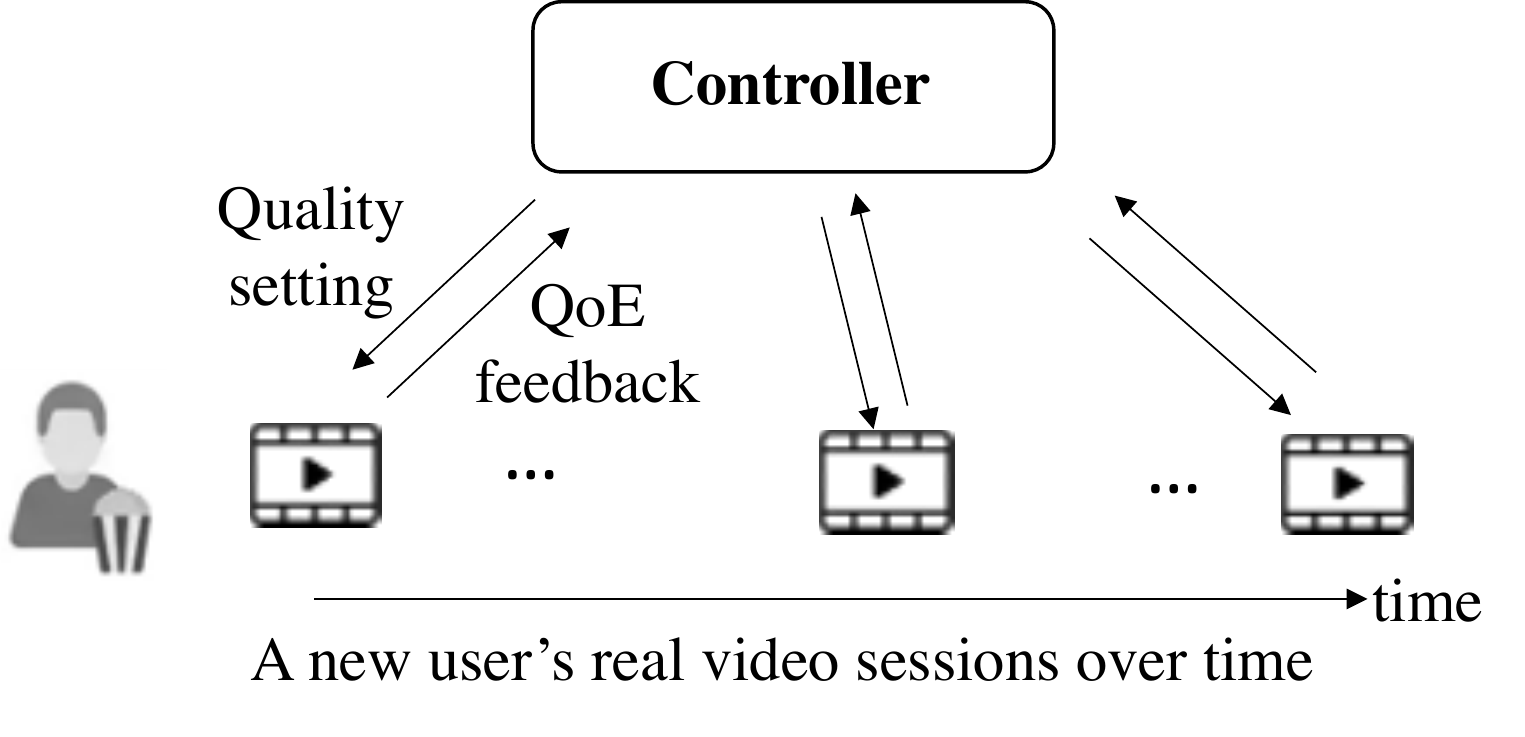}
    \caption{Framework of online QoE modeling and optimization.}
    \label{fig:online-framework}
\end{figure}

We envision online QoE modeling and optimization in terms of a simple framework that a service provider can use to profile per-use QoE models and then optimize QoE for new users.
Depicted in Figure~\ref{fig:online-framework}, this framework allows users to watch videos in their natural environments (any content users wish to watch at any time). 
It contains a logical controller (which could be operated by a service provider or by the player) that determines the video quality during each video session.
At the same time, the controller monitors the QoE (defined shortly) of each session and uses the measurements of QoE (subjective) and the video quality (objective) from each session to update the per-user QoE model maintained by the controller. 

\begin{table*}[]
\small
\begin{tabular}{ccccc}
\hline
 & \begin{tabular}[c]{@{}c@{}}{\bf R1:} No control on when \\ \& what users watch\end{tabular} & \begin{tabular}[c]{@{}c@{}}{\bf R2:} Fast (min \# of \\ profile videos)\end{tabular} & \begin{tabular}[c]{@{}c@{}}{\bf R3:} High QoE \\ during profiling\end{tabular} & \begin{tabular}[c]{@{}c@{}}{\bf R4:} Accurate per-user \\ QoE model\end{tabular} \\ \hline
Lab study~\cite{duanmu2019knowledge,duanmu2016sqi} &  &  &  & \ding{52} \\ \hline
Offline data analysis~\cite{gao2020personalized} & \ding{52} &  & \ding{52} & \ding{52} \\ \hline
Greedy profiling~\cite{chang2018active} & \ding{52} & \ding{52} &  & \ding{52} \\ \hline
Greedy optimization~\cite{menkovski2010online} & \ding{52} &  & \ding{52} &  \\ \hline
\name (ours) & \ding{52} & \ding{52} & \ding{52} & \ding{52} \\ \hline
\end{tabular}
\vspace{0.4cm}
\caption{Requirements of an ideal realization of online QoE modeling and optimization, and in what aspects prior solutions are ill-suited.}
\vspace{-1.7em}
\label{tab:comparison}
\end{table*}

In modeling the QoE preferences for each user, an ideal realization of an online QoE modeling and optimization framework should meet four properties (Table~\ref{tab:comparison}):
\begin{packeditemize}
\item {\bf R1:} It should not control when and what users watch, since real service providers (CDN, ISP) or content providers are unlikely to specify what videos real users must watch first. 
\item {\bf R2:} It should finish QoE modeling within a small number of video sessions, because service providers want to optimize each of their users as early as possible. 
\item {\bf R3:} It should provide high QoE while modeling per-user QoE, \ie one cannot model per-user QoE at the cost of user experience. 
\item {\bf R4:} Finally, we want the eventual per-user QoE models to be accurate enough for per-user optimization techniques (ABR logic and bandwidth allocation) to best leverage the heterogeneity across users.
\end{packeditemize}

\begin{figure}[t]
    \centering
    \includegraphics[width=0.7\linewidth]{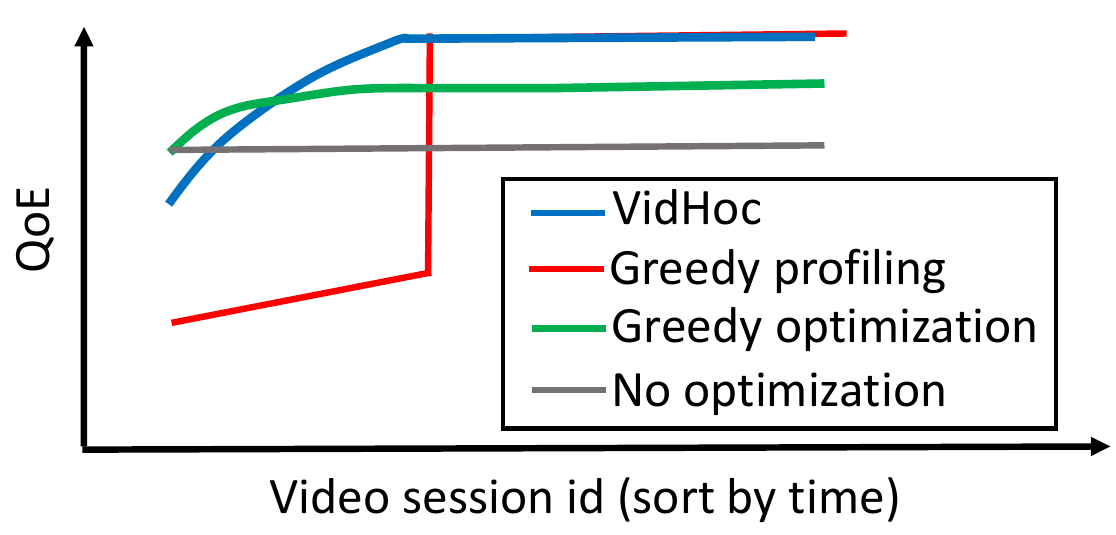}
    \caption{QoE over time of an ideal realization of online QoE modeling and optimization (blue) where QoE improves quickly as per-user QoE model becomes more accurate. 
    It outperforms the strawman of active-learning-based aggressive profiling (red) which uses low-quality videos to elicit QoE feedback) and greedily optimizing QoE (green) which relies on the latest though inaccurate per-user QoE model. 
    }
    \label{fig:ideal-timeseries}
\end{figure}

Figure~\ref{fig:ideal-timeseries} shows a desirable behavior of \name (blue line): 
For each user, the quality and QoE measurements of each video session will be used to update the user's QoE model, and over time, the QoE model will become more accurate such that future video sessions can leverage the latest per-user QoE model to improve quality over the baseline (grey) that is agnostic to per-user QoE preferences. 

\tightsubsection{Why is it different?}
\label{subsec:strawmen}

Table~\ref{tab:comparison} puts existing techniques into the context of these four dimensions. 

First, many previous QoE studies, based on controlled lab studies or crowdsourcing surveys~\cite{zhang2021sensei,duanmu2019knowledge,duanmu2016sqi}, are ill-suited, because they require users to watch videos of specific content while varying the video quality to induce QoE samples to build a good QoE model (\ie violating {\bf R1}).
Alternatively, offline data analysis is commonly employed by service providers who are in a position to passively collect massive data on quality and QoE of a large number of sessions from each user~\cite{zhao2019qoe}. 
While analyzing such large datasets might ultimately yield accurate QoE models, the QoE modeling is still done offline (\ie violating {\bf R2}), rather than as an integrated part of the online QoE optimization framework. 

To address these shortcomings, two seemingly natural designs emerge.
Both rely on passive QoE measurements and update the per-user QoE model after each video session.
They differ in whether they set video quality to maximally improve user QoE or maximally help QoE modeling.

\mypara{Greedy optimization using biased QoE models}
As more measurements from a user are collected to update the per-user QoE model, the player can use the latest per-user QoE model as input to the ABR algorithm, under the assumption that the latest QoE model is accurate enough.
This is promising, because the video quality would gradually improve as the QoE model becomes more accurate.
The drawback, however, is that the QoE model may never be accurate enough to realize the full potential of per-user QoE optimization.
This is because 
the latest QoE model may be inaccurate and thus introduces bias in the ABR decisions. 
For instance, 
YouTube players typically assume that rebuffering has a greater impact on QoE than low bitrate, so it strives to minimize rebuffering~\cite{anorga2015youtube} by aggressively lowering bitrate whenever bandwidth drops. 
However, as we have seen in \S\ref{sec:moti:hetero}, some users are actually more sensitive to lower birate than to rebuffering. 
So the QoE model may never learn how sensitive each user actually is to rebuffering, thus missing opportunities to optimize QoE (\ie violating {\bf R4}).

\mypara{Greedy profiling using active learning}
Another natural choice is to add quality issues to video sessions in a way that the user's QoE feedback will maximally reduce the uncertainty of the current QoE model. 
Previous works~\cite{chang2018active} show that active learning (which shows users the video quality whose QoE the current model is most uncertain about) can be used to minimize the number of sessions (QoE samples) needed to build an accurate QoE model;
\eg it only needs $\sim$100 videos to get the per-user QoE model from scratch.
This method is not ideal either. 
Although we can train per-user QoE models quickly and start QoE optimization early, the videos in the training stage might require severely bad quality (\ie violating {\bf R3}), because the video quality patterns (\ie a video quality pattern is the quality of each chunk in the video session.) with the largest uncertainty in the QoE model are usually with bad video quality.
To accelerate per-user QoE modeling, the active learning-based methods tend to encapsulate multiple types of quality issues into one video, and the low video quality can cause tens of billions of dollars profit damage according to Conviva~\cite{convivadamage}.


\tightsection{Design of \name}
\label{sec:design}

In this section, we describe a specific setting for which we would design a concrete solution, {\em \name}, to realize online QoE modeling and optimization.

\tightsubsection{Overview}

The design space of online QoE modeling can be characterized along two dimensions:
\begin{packeditemize}
\item {\em What QoE feedback} will be collected from each session?
\item {\em What control knobs} will be used to influence video quality? 
\end{packeditemize}

\mypara{Setting of \name}
\xu{In this paper, \name is a realization of online QoE modeling and optimization in a specific design point.
We put \name under the following practical settings and limitations.
First, we use {\em user engagement} as the metric of user QoE.
Although user engagement can be affected by the factors other than the video quality, \eg video content, there is still a strong correlation between user engagement and the video quality, especially for the users who suffer from the video quality issues.
Another reason of using user engagement is to minimize the user annoyance and the cost of collecting user QoE feedback.
A popular method is to solicit the user feedback through an exit survey after a video session ends.
We argue that it is either less incentive due to the user annoyance or costly if the service provider pays for the user feedback, which prevents it is large-scale deployment in the production systems.
Second, we do not control the video content or directly change the behavior of the video player.
Instead, \name passively monitors the player status (\eg video resolution, buffer length, etc) to infer the video quality which can be used to train per-user QoE model with user engagement.
It then dynamically changes the available bandwidth to indirectly control the video player's bitrate selection.
Third, other than the video quality, we do not interfere with the other experience of viewing videos (\ie context factors), including user interface (UI), player settings, etc., which can also affect user engagement.
Thus, when we train the QoE models, the video quality is the most significant factor of user engagement other than the video content.
}

This setting can be suitable for network service providers, such as ISPs, CDNs, and edge routers, who control the bottleneck bandwidth of a user. 
The network service providers do not have the access to the video player's logic but can monitor and control the users' network connection.
To obtain the video quality, service providers can use  existing techniques to infer video quality by the video streaming traffic \cite{bronzino2019inferring}, and use the session time of the video streaming as the user engagement, and can guess the player status with up to 91\% accuracy. 
We evaluate the impact of the player status inference accuracy and the QoE gain of \name in \S\ref{sec:eval}.
Our evaluation results show potential benefits to these service providers in practice: \name can improve QoE (user engagement) without changing the average bandwidth per session, or save bandwidth to achieve the same QoE (\ie if \name is deployed on the bottleneck link, it might support more concurrent users without degrading their QoE).

A natural concern regarding this setting is whether user engagement measurements from arbitrary video content are reliable enough to model per-user QoE.
Intuitively, video content does impact users' sensitivity to video quality.
However, in our study, the selection of quality level is totally agnostic to video content.
Thus, averaging over a sufficient amount of data can still reliably build per-user QoE models.
We make this as a pragmatic choice, because our user study cannot record video content.
That said, we hope future studies will take video content and other factors (\eg viewing device) into the modeling of per-user QoE.

\begin{figure}[t]
    \includegraphics[width=1.0\linewidth]{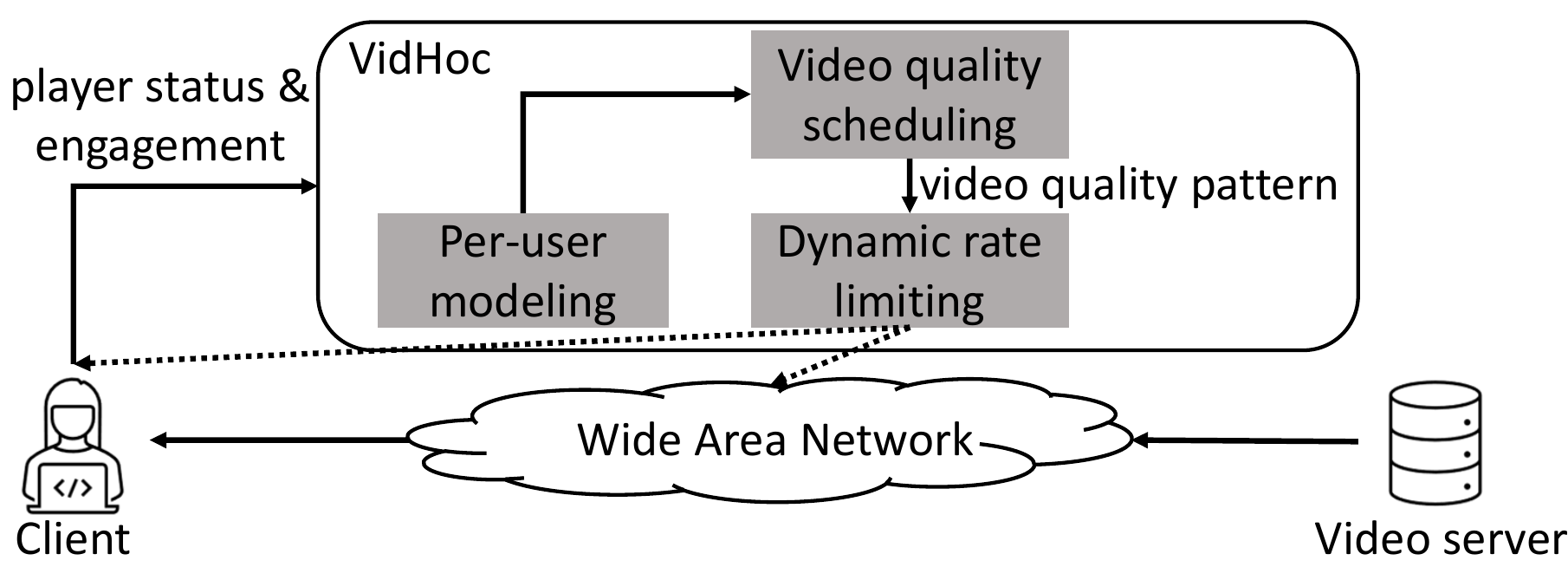}
    \caption{System overview of \name. 
    }
    \label{fig:overview}
\end{figure}

\mypara{Components} 
Based on the setting, we envision \name has three modules as shown in Figure~\ref{fig:overview}: 

\begin{packeditemize}
\item {\em Video quality scheduling:}
For each video session, \name first determines the {\em quality pattern} (\ie the quality of each video chunk) of the session.
Ideally, the quality pattern itself should induce decent QoE, while the QoE measurement from the chosen quality pattern would improve the accuracy of the QoE model. 

\item{\em Dynamic rate limiting:}
Once the video quality pattern is determined, we then need to decide how to dynamically set available bandwidth to realize the quality pattern.
Ideally, if the video players can rapidly adapt to the downloading rate, we can set the rate as the bitrate of the video quality we want.


\item{\em Per-user QoE modeling and optimization:}
Finally, the measurements of engagement and video quality from each user are fed to update the user's QoE model.
Since the design of the QoE model itself is not our contribution, we reuse a random-forest-based QoE model from prior work~\cite{chang2018active}
and use per-user QoE measurement data to train the model for each user.
\footnote{
In particular, the QoE model works as follows. 
This model takes the video quality as the input variables, and outputs the expected QoE as the output.
The input variables are the video quality of the chunks.
Based on the QoE model we currently have and the video trace, we can decide the video quality pattern to show the users.
We customize feature selection and criteria at the nodes on each decision tree inside the forest.
To train the QoE model faster, the initial QoE model is a one-size-fits all model.
To avoid the explosion of input due to the large number of video chunks, we divide the video into 20 parts, and QoE model will take the quality of each part.
The total number of input variables are 20 * 3 = 60.
The video quality of each part includes its average video bitrate, total bitrate switch, total rebuffering time.
The output is the user QoE.
The QoE is bucketized the user engagement into 10 classes, and the output is the class who gets the most votes from the decision trees inside the random forest.}

\end{packeditemize}

Next, we will explain how \name's quality scheduler and rate limiting module achieves meet the four requirements.

\tightsubsection{How \name schedules video quality}
\label{subsec:design:balancing}

The goal of \name is to model QoE online, \ie in real-time, while maintaining good video QoE.
The two solutions described in \S\ref{subsec:strawmen} are \xu{not ideal}:
greedy optimization leads to sub-optimal video quality in the long run since it relies on a biased QoE model, while
\xu{greed profiling severely impairs the video quality early on during QoE modeling, and it might discourage the users to use this video streaming service.}

To address this challenge, our insight is to cast the problem into a well-studied frame of regret minimization~\cite{loomes1982regret}.
For each video session $i$, we select a video quality pattern $v_i$ among those that are possible stream under a given average bandwidth limitation.
Mathematically, the picked quality pattern should maximize the sum of the estimated expectation of QoE and the uncertainty of this estimation (explained shortly).
Formally, we solve the following problem to select the quality pattern $v_i$:
\begin{align}
    \max& \quad\quad QoE(v_i)+Uncertainty(v_i), \label{eq:find_patter}\\
   \text{s.t.}&  \quad BandwidthUsage(v_i) < BandwidthLimit\label{eq:bt_limit},
\end{align}
$QoE(v_i)$ is the predicted QoE \xu{(\ie predicted user engagement here)} of the video quality pattern $v_i$, $Uncertainty(v_i)$ is the uncertainty of this prediction, the $BandwidthLimit$ is the pre-determined maximum average bandwidth per session (\eg this can be based on the Internet plan subscribed by the user), and $BandwidthUsage(v_i)$ is calculated by dividing the sum of bitrates of expected watched chunks (the ones within the predicted viewing time $QoE(v_i)*VideoLength$) by the predicted viewing time.

Given the current QoE model (a random forest of decision trees) and a quality pattern $v_i$, each decision tree returns a predicted QoE bucket (a range of user engagement), and we count the number of ``votes'' given by these decision trees to each possible QoE bucket. 
QoE prediction $QoE(v_i)$ is the median of the QoE bucket with the most votes.
The uncertainty $Uncertainty(v_i)$ of this prediction is measured by the difference between the highest vote count and the second highest vote count. 
(This way of measuring uncertainty is also called {\em minimal margin}~\cite{settles2009active}.)
Intuitively, the uncertainty of a video quality pattern indicates how much the current QoE model is unsure about predicting the QoE of this quality. 
The constraint of Eq.~(\ref{eq:bt_limit}) ensures that it is possible to stream the video under the quality pattern $v_i$ within the expected viewing time $QoE(v_i)$ without using more bandwidth than $BandwidthLimit$.

Once a session ends, a pair of its actual session-level quality and user QoE (user engagement) is used to update the current QoE model.
This can be done in various ways.
In our implementation, we train the initial QoE model of each user based on an ``initial dataset'' of 120 sessions which are collected from 10 users\footnote{Those users met the recruiting criteria in~\S\ref{sec:impl} but quit in the middle.} (different from the users whom we eventually test \name on) during the first 2 months of the user study.
And then the measurements (pair of session-level video quality and QoE) from a user will be added to a separate copy of the ``initial dataset''.
That is, each time a new measurement is collected from a user, it is added to the the user's version of  ``initial dataset'' and retrains the model for this user.\footnote{One needs to balance the data size of the ``initial dataset'' and the new samples from each user: if the initial dataset is too large, it will dominate the training of the per-user QoE model. In our implementation, the initial dataset (120 sessions) does not seem to overly dominate the per-user model, but we leave a thorough investigation of this issue to future work.}

An intuitive interpretation of Eq.~(\ref{eq:find_patter}) is that it minimizes the ``regret'' of choosing a quality pattern.
That is, we minimize 
(1) the suboptimality of the expected QoE (via a high $QoE(v_i)$), and 
(2) the impact on future QoE optimization (via choosing a quality that induces feedback on a highly uncertain QoE prediction $Uncertainty(v_i)$).
This formulation effectively balances the QoE of the current video session and the accuracy of QoE modeling which affects future QoE optimization---this balance is what the two strawmen discussed in \S\ref{subsec:strawmen} do not address.

\tightsubsection{Dynamic rate limiting}
\label{subsec:dynamicratelimiting}
In our setting, a practical concern is that we cannot directly control the player's selection of quality.
Instead, we control the downloading rate (\ie bandwidth) during the course of a session to indirectly control the quality pattern \xu{by inducing the ABR algorithm to adapt to the downloading rate we set}.
As a result, we might fail to achieve the video quality pattern selected by \name's quality pattern scheduler (\S\ref{subsec:design:balancing}).
Thus, we must take the player's behavior in bitrate selection into account.


First, numerous prior studies have shown that ABR-enabled players can adapt its video quality in response to changes in the available bandwidth (In fact this is used by several papers to reverse-engineer video player logic~\cite{gruner2020reconstructing}).
For example, if we want to the player to change its video bitrate from 1Mbps to 500Kbps at the $10^{\textrm{th}}$ chunk, we can lower the bandwidth from 1Mbps to 500Kbps when the player downloads the $10^{\textrm{th}}$ chunk.
An important takeaway from these prior works is that there is usually a non-trivial delay between when the bandwidth changes and when the player responds, and the bitrate which the player switches to is often a distribution, rather a fixed value. 

This observation inspires our solution: 
instead of determining how to induce the player to {\em deterministically} change its bitrate, a more pragmatic approach would be to model {\em distribution} of player's action (when to switch to what bitrate).
In our implementation, we first obtain this distribution on the YouTube player\footnote{YouTube is known to occasionally change the player ABR logic. In this test, we use the YouTube player obtained during Sept 2021 - Dec 2021.} that we test in our user study as follows. 
We run an exhaustive test to build the transition distribution of $D(s_1, s_2, b)$, where $s_1$ and $s_2$ represents the player states, including current bitrate, buffer length, and $b$ is the current bandwidth.
To obtain the transition probability between $s_1, s_2$ under bandwidth $b$ within 12 seconds\footnote{We use 12 seconds because empirically it is the medium delay between when bandwidth changes and when the video player reacts.}, we set the player state to $s_1$ under bandwidth $b_1$ for 100 times, and $D(s_1, s_2, b)$ is the times of the states becoming $s_2$ divided by 100.
Due to the infinite number of all possible player states and bandwidth, we bucketize the bitrate, buffer length and bandwidth for every 200Kbps, 1 second and 200Kbps respectively.
These transition distributions are inherently specific to the video player under study, and we do not claim they are generalizable to other players.

The resulting player's transition distributions fit nicely to the formulation of Eq.~(\ref{eq:find_patter}).
Instead of finding a quality pattern $v_i$ for session $i$, our optimization objective is to find a rate-limiting scheme $b_i$ (bandwidth over time) that can minimize the {\em expected} regret of the video session:
\begin{align}
  \min& -\sum_{v_i}p\left(b_i, v_i\right)*[QoE\left(v_i\right)+Uncertainty\left(v_i\right)], \label{eq:bw_regret}\\
  \text{s.t.}& \quad BandwidthUsage(b_i)\leq BandwidthLimit,
\end{align}
where $p(b,v)$ is the probability that the video quality pattern $v$ appears under the rate-limiting scheme $b$, and $p(b,v)$ can be calculated by the player profile by going through all possible video quality patterns.
During a video session, the player status is updating over time, and $p(b,v)$ updates accordingly.
Thus, we can update the rate-limiting scheme $b$ by the updated $p(b,v)$.

Note that there are multiple possible video quality patterns that can be generated by the same bandwidth allocation plan, and we might not get the video quality pattern with the minimal regret.
We admit it could happen, but we can minimize the negative impact, \ie little contribution to the QoE model training and/or bad user QoE.
First, we already consider the possibility of getting the other video quality patterns, since those video patterns' regrets are calculated in the regret expectation of the bandwidth allocation plan.
Second, one might argue that we might calculate the wrong expectation because the profile of the player behavior might not be accurate.
It does happens if the ABR logic changes or the player behavior profile has insufficient data.
Thus, we can either re-profile the player's behavior or add more data points to the existing profile.
Third, even if the worst case happens, its negative effect is still small.
The goal of \name is to maximize both QoE and uncertainty.
If the resulted video quality pattern has small uncertainty, it is still better that GreedyOpt that does not interfere with the video quality at all.
Because, \name creates more bandwidth fluctuations, and thus there would be more quality issues than GreedyOpt in the video quality patterns, which also leads to a faster training speed than GreedyOpt.
If the resulted pattern has the bad quality, it still has better quality than GreedyOpt.
It happens when we expect a slightly lower quality but get a much worse quality level, \eg we expect a small bitrate degradation but get a large bitrate drop or even a rebuffering event.
In this case, \name could re-calculate the bandwidth allocation plan and allocate more bandwidth to the following chunks for a better regret, and the more bandwidth will lead to a better video quality.
One might concern that the player might not increase the bitrate adapting to the increased bandwidth.
Fortunately, it rarely happens because the goal of the players is to maximize QoE in which bitrate is an important factor.
The previous work \cite{?} (Understanding video streaming algorithms in the wild
) shows that although some ABR algorithms used in industry have a conservative choice on bitrate selection, it will still increase the bitrate for adapting to the bandwidth.

\tightsubsection{Reducing controller compute overhead}
\label{subsec:costsaving}
To naively calculate the expected regret of a rate-limiting scheme (Eq.~(\ref{eq:bw_regret})), we need to enumerate all possible video quality patterns under this scheme, which can be prohibitively expensive.
We reduce this overhead by two ideas that limit the number of patterns we need to enumerate.

First, we use a short look-ahead horizon, such that we do not need to enumerate all possible quality of the future chunks.
For the later chunks, we assume the player will perfectly adapt to the bandwidth we set.
In \name, the look-ahead horizon is 2, and in \S\ref{subsec:microbenchmark}, we will evaluate the impact of different look-ahead horizons on QoE improvement and compute overhead.

Second, we coalesce multiple consecutive chunks as longer segments, so that the bandwidth scheme and quality patterns are determined over a handle of segments, rather than many more chunks. 
For example, for a 60-second long video with 3-second chunk size, we need to enumerate 60/3=20 times, while if we consider 4 adjacent chunks as a whole, we only need to enumerate 60/(3*4) = 5 times, which saves more than 400\% computation cost.

\tightsection{User study methodology}
\label{sec:impl}


In this section, we present the user study methodology to evaluate \name and the baselines.



\begin{figure}[t]
	\centering
	\includegraphics[width=1.0\linewidth]{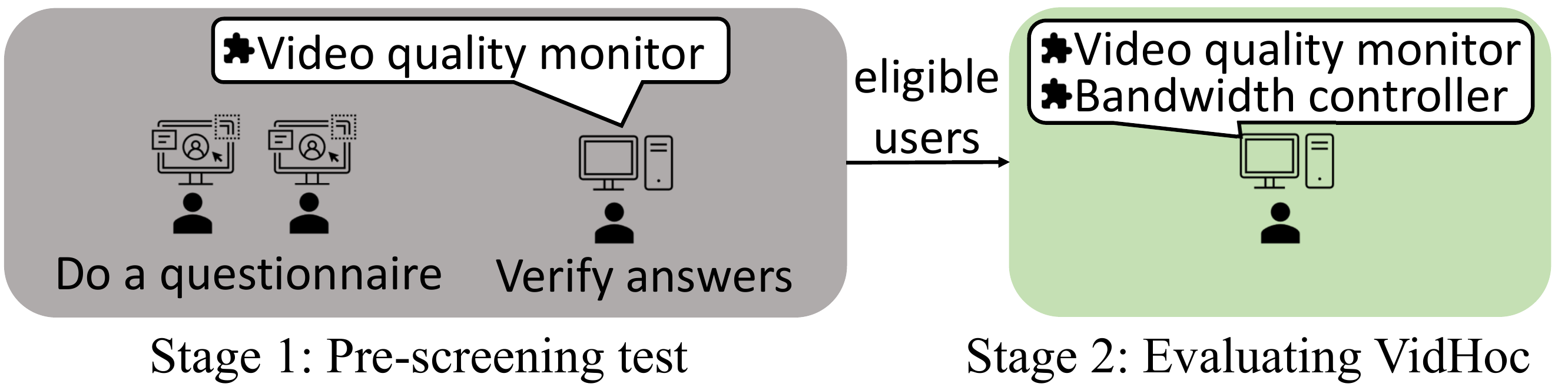}
	\caption{User study procedure: We first send out the filtering survey to crowd workers and verify their answers by monitoring their video quality. Then the eligible users are asked to install a browser extension that monitors video quality and controls throughput of YouTube sessions, which allows us to test \name and the baselines.}
	\label{fig:user_study_overview}	
	\vspace{-0.6em}
\end{figure}

We test \name and the baselines on the users recruited from two popular crowdsourcing platforms, \ie Amazon Mechanical Turk~\cite{mturklink} and Prolific~\cite{prolificlink}.
Figure~\ref{fig:user_study_overview} shows the two-stage procedure of conducting user study to evaluating \name on real users. Our study was approved by our Institutional Review Board (IRB).

First, to recruit the participants in our study, we published a filtering survey on two crowdsourcing platforms to solicit background and demographic information including age, gender, internet plans, frequency of watching online videos, and frequency of suffering quality issues from  potential participants.
Since we want to recruit users with room to improve their QoE, we selected eligible participants 
who used video streaming services more than $\geq$ 10 hours per week and who suffered from quality issues frequently ($\geq$20 times per week).
We restricted to participants living in the US.
To evaluate \name on different populations, we recruited a diverse set of participants' in terms of gender and age as shown in Table~\ref{tab:userstats}, which are similar to the demographics of YouTube users~\cite{youtubestats}.


After filtering, we asked selected users to install a browser extension implementing \name.
Our browser extension can limit the throughput of YouTube video sessions, and passively record the metrics of user's video sessions (no video content or private information is recorded), including video quality (resolution, rebuffering events), player status (player progress slider, buffer length, etc.) and video session time.
Unlike the previous similar study that asks for user feedback with an exit survey after every video session ends (\eg~\cite{zhang2021sensei}), our browser extension only monitors view time (or user engagement) as the QoE metric, which minimizes the user annoyance.

\begin{table}[t]
\begin{tabular}{|ll||ll||ll|}
\hline
Age      & \# & Gender & \# & Internet plan & \# \\
\hline
21 to 29 & 4  & Female & 7  & 25Mbps                                                                      & 12 \\
30 to 39 & 8  & Male   & 8  & 50Mbps                                                                      & 3  \\
40+      & 3  &        &    &                                                                             &   \\
\hline
\end{tabular}
\vspace{1em}
\caption{User Demographics}
\vspace{-2em}
\label{tab:userstats}
\end{table}

We conducted this user study from September to December 2021 on 6 users from Amazon Mechanical Turk and 9 users from Prolific.
In the pre-screening stage, we paid participants \$10 for completing the filtering survey and installing the browser extension in accordance with the minimum wage in ANONYMIZED.  
Participants were also paid an additional \$2 a week for each week they remained in the study with our browser extension running.
The implementation details of the browser extension are described in Appendix~\ref{sec:impl:plugin}.

\mypara{How to compare different schemes}
The practical challenge facing our evaluation is how to test different bandwidth control schemes on real users while we do not actually control the bottleneck bandwidth. 
\xu{First, for different schemes, the users watched videos under the same context (including viewing device, player display size, etc.) to make a fair comparison.
}
Moreover, we had to fairly compare competing schemes even though we do not control when users watch videos (\eg we cannot schedule users to watch $n$ videos to test method A and then another $n$ videos to test method B).
We address this issue by two simple-yet-practical ideas:

{\em Throughput capping:} During our user study, we use the browser extension to impose a fixed limit on the throughput of all video session, such that when \name and its baselines set bandwidth, the average throughput must not exceed $\alpha\%$ (we use 60\% in our study) of the actual throughput.
This way, all schemes will be subject to the same $BandwidthLimit$ (see Eq.~(\ref{eq:bt_limit})) of $\alpha\%$ of the actual throughput. At the same time, they still have the flexibility to decide what bandwidth schedule to use in each session.

{\em Randomized A/B tests:}
In addition, to mitigate the influence from video content (which are unknown to us), we borrow the idea from A/B test~\cite{young2014improving}: we randomly determine which strategy we use to optimize QoE when each video session begins.

\tightsection{Evaluation}
\label{sec:eval}

Using a randomized A/B tests on 15 real participants, our evaluation of \name shows the following key findings:
\begin{packeditemize}
\item When keeping the same average bandwidth per session, \name achieves better user engagement,
by 9.2\% compared to ``greedy optimization'' which does not customize for per-user QoE model, by 21.6\% compared to ``greedy profiling'' which greedily utilizes the latest (though inaccurate) per-user QoE model.
\item \name achieves similar QoE compared to various baselines but using much less average bandwidth per session: 
21.2\% less compared to ``greedy profiling'', 17.3\% less compared to ``greedy optimization'', and 20.7\% less compared to ``no optimization''. 
\end{packeditemize}


\tightsubsection{Setup}

\mypara{Baselines}
We compare \name with three baseline strategies (described in \S\ref{subsec:strawmen}):
\begin{packeditemize}
    \item {\em Greedy QoE optimization (GreedyOpt):} This strategy greedily optimizes user QoE by exploiting the QoE model that is updated after each video session.
    GreedyOpt does not deliberately add quality issues to the video sessions, but may suffer from inaccuracy of the QoE model.
    \item {\em Greedy QoE profiling (GreedyPF):}
    During the first 30 sessions (we will test other choices in \S\ref{subsec:microbenchmark}) of each user, this strategy adds quality issues to the video sessions to maximally and quickly reduce the uncertainty of the current QoE model, and it runs the same optimization as GreedyOpt. 
    \item {\em No optimization (NoOpt):} We do not have any intervention on the video quality. This represents the default ABR performance.
\end{packeditemize}

\mypara{Test methodology}
As described in \S\ref{sec:impl}, for a fair comparison, when a video session begins, we randomly determine which strategy is going to be used for this session.
After each session, we record video engagement (\ie the ratio of video session time and video length) as the user QoE.
We compare \name with other baselines in terms of the average QoE across the users under a fixed average bandwidth throttling rate of 60\% (\S\ref{sec:impl}). 

\begin{figure}[t]
    \centering
    \includegraphics[width=0.8\linewidth]{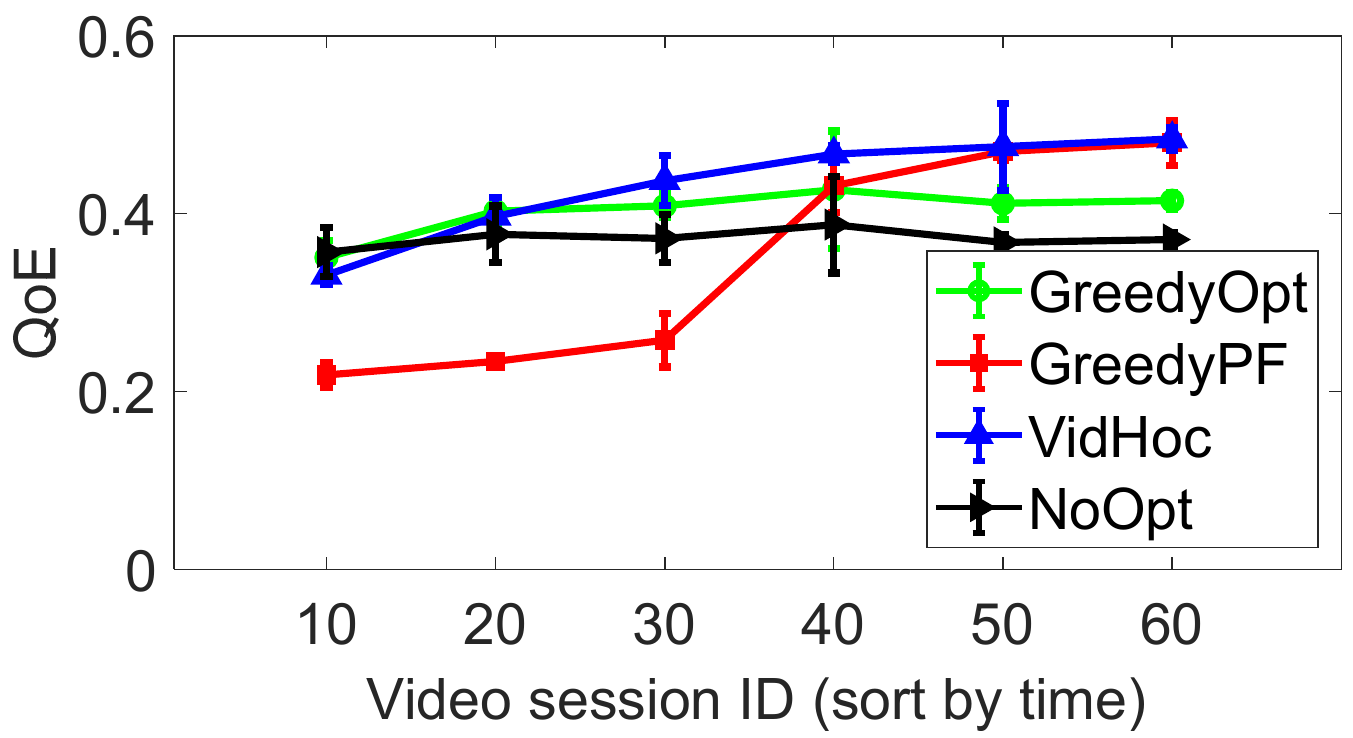}
    \caption{Improvement of average QoE over time across users of \name and other baselines.}
    \label{fig:qoe_gain}
\end{figure}

\begin{figure}[t]
    \centering
    \includegraphics[width=0.8\linewidth]{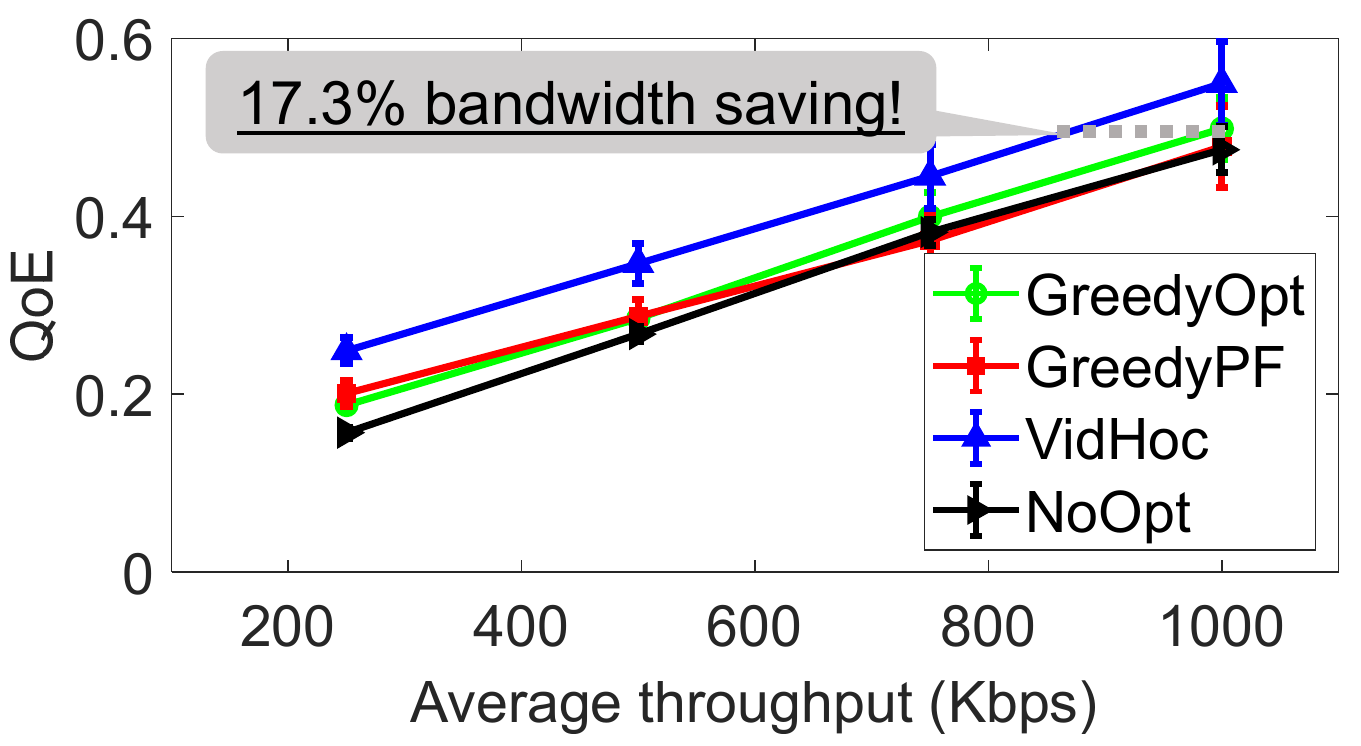}
    \caption{Average QoE across users vs. average throughput buckets during the first 60 sessions of each user.}
    \label{fig:bw_saving}
\end{figure}

To complement the real-world tests, we also leverage the large amount of traces collected from the participants to create a trace-driven simulator.
The simulator helps us answer ``what-if'' questions (\eg how much would the improvements be if \name uses a different parameter) to understand how sensitive \name's performance is to its parameters. 
The simulator uses all data collected from each user to build the per-user QoE models and uses them to simulate user QoE when \name (and baselines) chooses different actions (the details of trace-driven simulator are in Appendix~\ref{sec:simulator}).
As the simulator embeds some necessary assumptions, we present the overall performance comparisons between \name and its baselines using real-world tests (Figure~\ref{fig:qoe_gain} and Figure~\ref{fig:bw_saving}), rather than simulation.
\xu{Throughout the period of user study, every user watched videos under the same individual context: the same viewing device, volumn, and the actual display size of the video  player.}

\begin{figure*}[t]
\begin{minipage}{\linewidth}
      \centering
      \begin{minipage}{0.22\linewidth}
          \begin{figure}[H]
              \includegraphics[width=\linewidth]{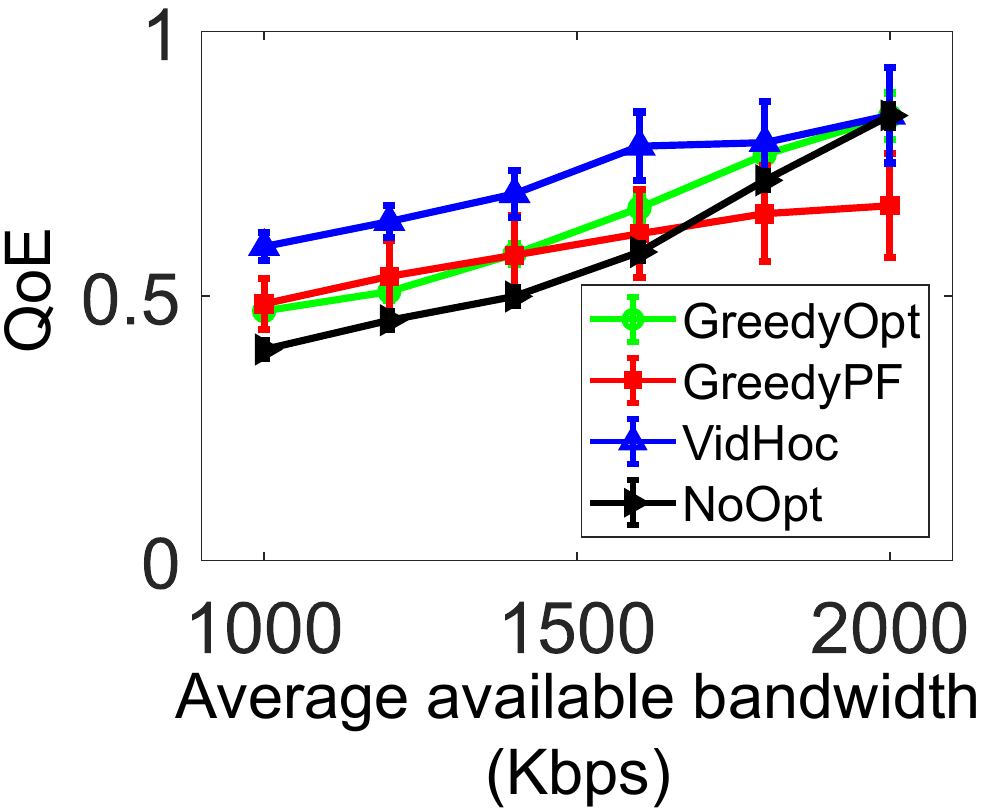}
              \caption{Average QoE across users in a wider (simulated) range of bandwidth. }
              \label{fig:bw_saving_project}
          \end{figure}
      \end{minipage}
      \hspace{0.01\linewidth}
      \begin{minipage}{0.22\linewidth}
          \begin{figure}[H]
              \includegraphics[width=\linewidth]{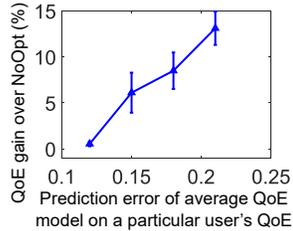}
              \caption{QoE gains over NoOpt vs. how much a user's QoE model is different from the average QoE model.}
              \label{fig:model_difference}
          \end{figure}
      \end{minipage}
      \hspace{0.04\linewidth}
      \begin{minipage}{0.48\linewidth}
          \begin{figure}[H]
              \includegraphics[width=\linewidth]{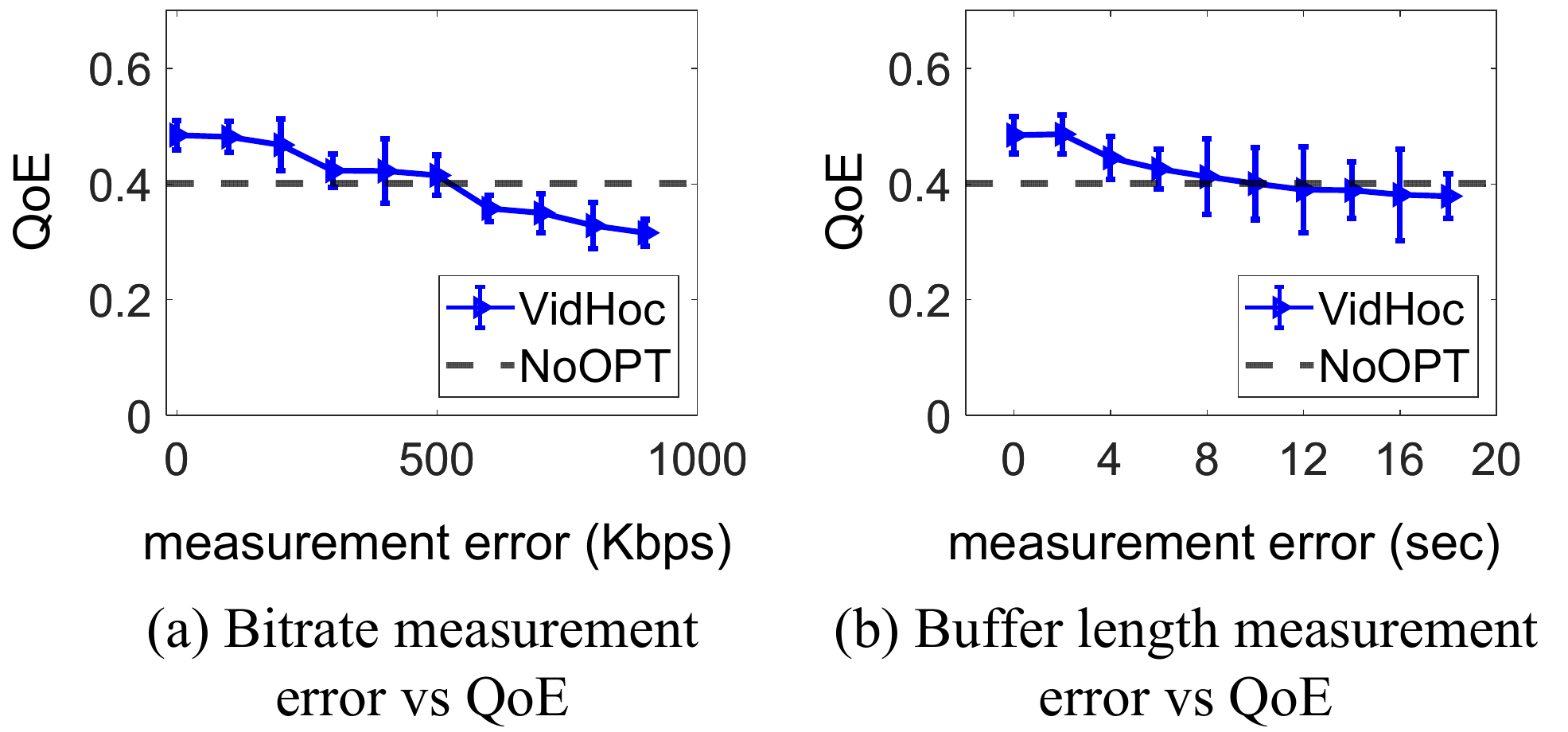}
              \caption{QoE under higher \name's measurement error on the player status.}
              \label{fig:guess_error}
          \end{figure}
      \end{minipage}
\end{minipage}
\end{figure*}

\tightsubsection{Overall performance}

\mypara{QoE improvement}
Figure~\ref{fig:qoe_gain} shows how average QoE of different schemes improves as a user watches more videos, when all schemes keep their average bandwidth usage at 60\% of the actual throughput.
This confirms the expected behavior outlined in Figure~\ref{fig:ideal-timeseries}: \name can quickly build a reliable per-user QoE model with a marginal QoE degradation in the first few sessions after a new user joins the service.
For the first 60 sessions, \name achieves on average $9.2-21.6\%$ higher QoE than the other baselines under the same average bandwidth limit. 
GreedyPF has much worse QoE than other methods during first 30 sessions when it builds the QoE model, but after that, it has the similar QoE as \name.
The video QoE of GreedyOpt is not as high as \name, since it uses sub-optimital QoE model to optimize user QoE.

\mypara{Bandwidth saving}
Next, we run the same real-world test in Figure~\ref{fig:qoe_gain} but with different limits on average bandwidth usage.
Figure~\ref{fig:bw_saving} shows the average bandwidths and the corresponding average QoE of different schemes during the first 60 video sessions after these schemes take effect. 
We can see that \name can save $17.3-21.2\%$ bandwidth while achieving similar QoE compared to other baselines.
At the same time, we notice that under low the available bandwidth (\eg $250 Kbps$), GreedyPF achieves similar QoE as \name and GreedyOpt, because such low bandwidth would naturally create various types of quality issues in the video quality patterns, so GreedyPF does not pay extra QoE penalty to profile per-user QoE models.

\tightsubsection{What affects \name's improvements}

We investigate the operational regime of \name, \ie when \name has QoE gain over the other baselines.

\mypara{User available bandwidth}
We have seen that \name can improve the QoE of the users who frequently suffer from quality issues in the video streaming service (\S\ref{sec:impl}).
However, intuitively, if a user already has a stable network connection that can sustain highest bitrate without rebuffering, there would be little room to improve this user's QoE. 
To validate it, 
we scale up the bandwidth of the video sessions in Figure~\ref{fig:qoe_gain} to $[1000, 1200, 1400, 1600, 1800, 2000]Kbps$ and simulate the user QoE by the per-user QoE model we obtained in the experiment in Figure~\ref{fig:qoe_gain}.
Figure~\ref{fig:bw_saving_project} shows our QoE improvement over other baselines diminishes with higher available bandwidth during the first 60 video sessions.
Indeed, \name has diminishing gains when the available bandwidth is large, because 1) the users rarely suffer from quality issues, and 2) \name still needs to deliberately show the videos with quality issues for per-user QoE modeling.

\mypara{Difference from the average QoE model}
Another factor that influences \name's gains is how different a new user's true per-user QoE model is to the average QoE model, based on which \name customizes the QoE model per user. 
Figure~\ref{fig:model_difference} shows \name's QoE gain over the baseline of GreedyOpt increases when the per-user QoE models more different from the average QoE model (\ie higher prediction error by the average QoE model).
On the other hand, if the average QoE model is very similar to the true per-user QoE model (\ie lower prediction error), GreedyOpt will become the optimal strategy, because we do not need to further learn the user's QoE model by adding quality issues.

\mypara{Player status estimation error}
In practice, we might not know the exact player status (\ie video bitrate and player buffer length), and need to infer the status for dynamic rate limiting.
In Figure~\ref{fig:guess_error}, we evaluate the robustness of \name when player status is estimated with errors.
We simulate this effect by adding white Gaussian noise to either bitrate or buffer length estimation.
We can see that \name's QoE might even fall below NoOpt when the player guess error is too large (600Kbps could be the bitrate gap between 3 resolution levels~\cite{mao2017neural,yin2015control}).
However, fortunately, prior work~\cite{bronzino2019inferring} has shown that the estimation error over 3 resolution levels is relatively rare (less than $40\%$ chance to happen, \eg under [0, 50]Mbps network bandwidth, the probability of the resolution in $\{240, 360, 480\}$P is $>60\%$ for Netflix, YouTube, Amazon and Twitch video players).

\tightsubsection{Microbenchmarks}
\label{subsec:microbenchmark}
Finally, we microbenchmark \name's performance if we change its key parameters.

\begin{figure*}[t]
\begin{minipage}{\linewidth}
      \centering
      \begin{minipage}{0.66\linewidth}
          \begin{figure}[H]
              \includegraphics[width=\linewidth]{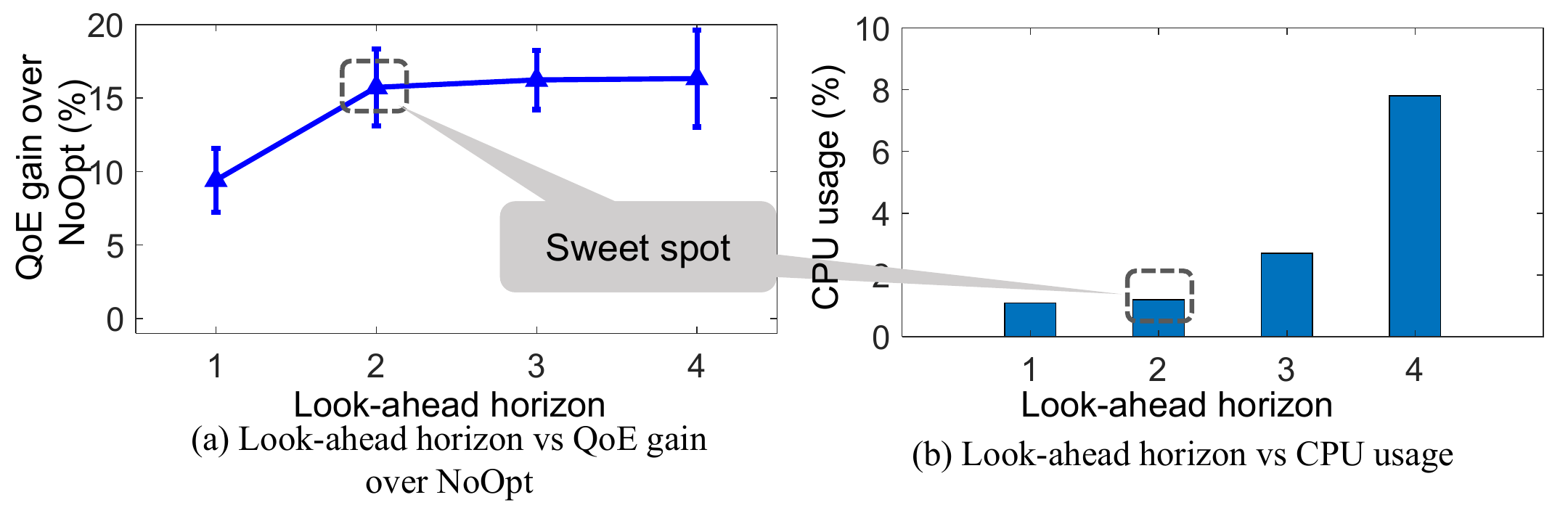}
              \caption{QoE and CPU usage under different look-ahead horizons.}
              \label{fig:look_horizon}
          \end{figure}
      \end{minipage}
      \hspace{0.002\linewidth}
      \begin{minipage}{0.3\linewidth}
          \begin{figure}[H]
              \includegraphics[width=\linewidth]{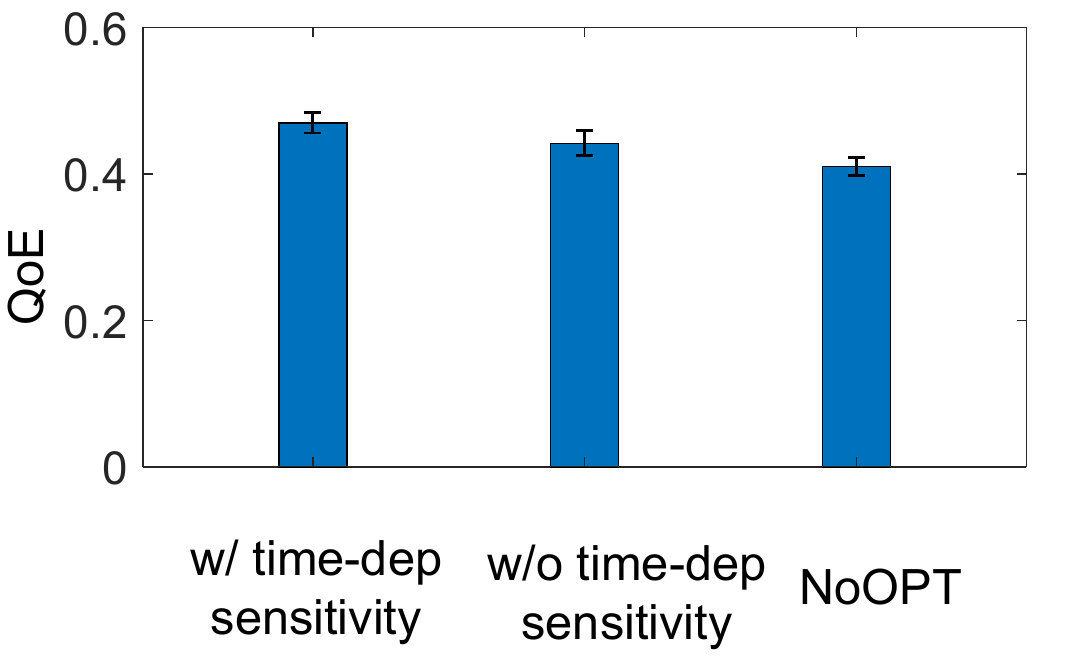}
              \caption{QoE improvement with and without taking time-dependent sensitivity into account.}
              \label{fig:time_sensi}
          \end{figure}
      \end{minipage}
\end{minipage}
\end{figure*}

\begin{figure}[t]
    \centering
    \includegraphics[width=0.7\linewidth]{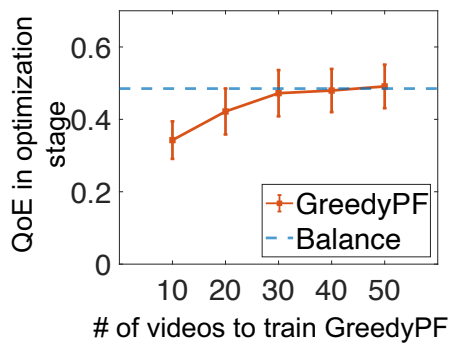}
    \caption{GreedyPF's QoE vs. \# videos for training.}
    \label{fig:al_parameter}
\end{figure}

\mypara{Look-ahead horizon}
In~\S\ref{subsec:costsaving}, we show that the computation cost is prohibitive if we enumerate all possible video quality patterns.
Figure~\ref{fig:look_horizon} shows the QoE gain over no optimization changes with the number of the future chunks that we enumerate their quality.
To balance the computation cost and QoE, we choose 2 as the look-ahead horizon, and
\name needs $\sim$250MB memory and 1.2\% CPU on a Debian 10 machine with Intel 3.2GHz CPU.

\mypara{Time-dependent QoE sensitivity}
Figure~\ref{fig:time_sensi} shows the QoE difference of \name whether we consider time-dependent sensitivity (which is shown in Figure~\ref{fig:memory}).
We can see that QoE can be improved by 6.4\% if we consider the time-dependent sensitivity, because it enables \name to have a time dimension to schedule bandwidth instead of the overall QoS metrics (\ie bitrate, rebuffering time, and bitrate switch).

\mypara{GreedyPF's training set size}
If we can use fewer video sessions to train the per-user QoE model by GreedyPF, we might mitigate GreedyPF's severe QoE degradation.
Figure~\ref{fig:al_parameter} shows the QoE in the optimization stage when we vary the number of videos for QoE modeling by deploying GreedyPF to real users.
We can use 30 videos to do QoE modeling for obtaining high QoE in the optimization stage while degrade less QoE during QoE modeling.


\tightsection{Related Work}

\mypara{User QoE modeling}
Past work \cite{bampis2018towards,bampis2017qoe,dobrian2011understanding,duanmu2019sqoe4,duanmu2017qoe,duanmu2018quality,duanmu2016sqi,grafl2013representation,tobias2013youtube,tobias2011youtube,ma2019gmad,moorthy2012video,ni2011flicker,pastrana2004sporadic,qi2006effect,rehman2013perceptual,staelens2010assessing} has studied the influencing factors on QoE and evaluated video streaming systems by conducting subjective user studies.
In those user studies, the users usually need to fill out an exit survey to provide feedback or their opinions on the video quality when a video session ends.
The user feedback is regarded as the end product of the ground-truth QoE and a zero-mean white Guassian noise~\cite{bt500subjective}.
Following this paradigm, previous research work focuses on predicting user QoE with perceptually relevant system QoS (quality of service) features including average bitrate, rebuffering duration, and quality adaptation magnitude~\cite{liu2012case,mok2012qdash,spiteri2016bola,yin2015control, bentaleb2016sdndash,duanmu2016sqi,duanmu2019knowledge,bampis2017atlas,itu2017pnats}.

Recently, many research studies focus on modeling user QoE in a finer granularity.
The work by \cite{zhang2021sensei} studies the QoE impacted caused by the video content, and \cite{gao2020personalized} reveals that the context of video viewing (\eg room brightness, viewing devices) is also important to user QoE.
The work by \cite{wang2016data} builds the QoE models for the individual users regarding the differences in viewing devices and points of interests.
To save the cost of exit surveys, some prior work minimizes the number of the video sessions that need user feedback by clustering users~\cite{gao2020personalized} or active learning~\cite{chang2018active}.
In this paper, \name further reduces the cost by passively monitoring the user engagement instead of directly asking for the user feedback, because of the strong correlation between user engagement and user's QoE rating~\cite{zhao2019qoe}.

\mypara{QoE-driven video streaming systems}
Early QoE-driven video streaming systems~~\cite{gomez2013towards,jiang2012qoe,ganjam2015c3,liu2012case,chiariotti2016online,mao2017neural,mok2012qdash,spiteri2016bola,yin2015control} adopt a one-size-fits-all optimization criterion, regardless of the video source characteristics, video viewing context, and individual users' quality preferences.
These algorithms are typically optimized offline on a dedicated server and then distributed to the operating point where resource allocation is performed.
Further research efforts mine information about video content~\cite{de2016complexity,toni2015optimal}, device type~\cite{bentaleb2016sdndash,li2016VMAF,rehman2015ssimplus}, and connectivity~\cite{jiang2016cfa,jiang2017pytheas} to optimize QoE at a per-group granularity.
Only until recently does personalized QoE optimization become a hot research topic.
Huo~\textit{et al.}~\cite{huo2020meta} discover the heterogeneity of QoE, manifested as the difference in the viewers' sensitivity to video quality, rebuffering, and quality adaptation.
Building upon the insights, the authors developed a data-driven personalized QoE model and a meta-learning optimization scheme.
As a few individual ratings become available, an optimal adaptive bitrate rule can be quickly determined with respect to the customized QoE model.
An alternative approach~\cite{gao2020personalized,wang2016data} takes user profile as the input feature to the resource allocation controller, effectively adapting to the individual characteristics at run time.




\section{Conclusion}
This paper fills a critical piece of video streaming systems that have been missing insofar: how to test the effectiveness of QoE customization schemes for new users?
We design and implement \name that can unbiasedly compare QoE customization schemes under real-world settings.
As a case study, we use \name to evaluate two natural and one improved QoE customization schemes on real human users.
The results reveal the potential of leveraging per-user QoE models: we can further improve QoE by 13.9\% with the same bandwidth or save 17.3\% bandwidth while maintaining the same QoE on the top of these natural schemes.

\mypara{Ethics Considerations} The experiments conducted in this paper were
conducted in accordance with accepted principles from the Menlo and Belmont
Reports. In particular, the Belmont Report speaks of respect for humans
(including informed consent), beneficence, and justice. All users for this
study were recruited and compensated in accordance with an experimental
protocol approved by our institution's IRB; users were
apprised of the experiment through an approved consent protocol and
compensated in accordance with minimum wage hourly rates. In this case, risks
to users were minimal (beneficence), and the user population that
participated in the experiment is of a similar demographic as that which
stands to benefit from this work (justice). Users were shown videos via
approved APIs and interfaces, thereby complying with the Menlo Report's
recommendation of respect for law and public interest.

\bibliographystyle{ACM-Reference-Format}
\bibliography{references}


\begin{thebibliography}{61}


\ifx \showCODEN    \undefined \def \showCODEN     #1{\unskip}     \fi
\ifx \showDOI      \undefined \def \showDOI       #1{#1}\fi
\ifx \showISBNx    \undefined \def \showISBNx     #1{\unskip}     \fi
\ifx \showISBNxiii \undefined \def \showISBNxiii  #1{\unskip}     \fi
\ifx \showISSN     \undefined \def \showISSN      #1{\unskip}     \fi
\ifx \showLCCN     \undefined \def \showLCCN      #1{\unskip}     \fi
\ifx \shownote     \undefined \def \shownote      #1{#1}          \fi
\ifx \showarticletitle \undefined \def \showarticletitle #1{#1}   \fi
\ifx \showURL      \undefined \def \showURL       {\relax}        \fi
\providecommand\bibfield[2]{#2}
\providecommand\bibinfo[2]{#2}
\providecommand\natexlab[1]{#1}
\providecommand\showeprint[2][]{arXiv:#2}

\bibitem[Amazon(2022)]%
        {mturklink}
\bibfield{author}{\bibinfo{person}{Amazon}.} \bibinfo{year}{2022}\natexlab{}.
\newblock \bibinfo{title}{Amazon Mechanical Turk}.
\newblock \bibinfo{howpublished}{\url{https://www.mturk.com/}}.
\newblock


\bibitem[A{\~n}orga et~al\mbox{.}(2015)]%
        {anorga2015youtube}
\bibfield{author}{\bibinfo{person}{Javier A{\~n}orga}, \bibinfo{person}{Saioa
  Arrizabalaga}, \bibinfo{person}{Beatriz Sedano}, \bibinfo{person}{Maykel
  Alonso-Arce}, {and} \bibinfo{person}{Jaizki Mendizabal}.}
  \bibinfo{year}{2015}\natexlab{}.
\newblock \showarticletitle{YouTube’s DASH implementation analysis}. In
  \bibinfo{booktitle}{\emph{19th International Conference on Circuits, Systems,
  Communications and Computers (CSCC)}}. \bibinfo{pages}{61--66}.
\newblock


\bibitem[Balachandran et~al\mbox{.}(2012)]%
        {balachandran2012quest}
\bibfield{author}{\bibinfo{person}{A. Balachandran}, \bibinfo{person}{V.
  Sekar}, \bibinfo{person}{A. Akella}, \bibinfo{person}{S. Seshan},
  \bibinfo{person}{I. Stoica}, {and} \bibinfo{person}{H. Zhang}.}
  \bibinfo{year}{2012}\natexlab{}.
\newblock \showarticletitle{A quest for an internet video quality-of-experience
  metric}. In \bibinfo{booktitle}{\emph{HotNets}}. \bibinfo{address}{Redmond,
  WA}.
\newblock


\bibitem[Bampis and Bovik(2017)]%
        {bampis2017atlas}
\bibfield{author}{\bibinfo{person}{C.~G. Bampis} {and} \bibinfo{person}{A.~C.
  Bovik}.} \bibinfo{year}{2017}\natexlab{}.
\newblock \showarticletitle{Learning to predict streaming video {QoE}:
  {D}istortions, rebuffering and memory}.
\newblock \bibinfo{journal}{\emph{ArXiv preprint arXiv:1703.00633}}
  (\bibinfo{date}{Mar.} \bibinfo{year}{2017}).
\newblock


\bibitem[Bampis et~al\mbox{.}(2018)]%
        {bampis2018towards}
\bibfield{author}{\bibinfo{person}{C.~G. Bampis}, \bibinfo{person}{Z. Li},
  \bibinfo{person}{I. Katsavounidis}, \bibinfo{person}{T.~Y. Huang},
  \bibinfo{person}{C. Ekanadham}, {and} \bibinfo{person}{A.~C. Bovik}.}
  \bibinfo{year}{2018}\natexlab{}.
\newblock \showarticletitle{Towards perceptually optimized end-to-end adaptive
  video streaming}.
\newblock \bibinfo{journal}{\emph{ArXiv preprint arXiv:1808.03898}}
  (\bibinfo{date}{Aug.} \bibinfo{year}{2018}).
\newblock


\bibitem[Bampis et~al\mbox{.}(2017)]%
        {bampis2017qoe}
\bibfield{author}{\bibinfo{person}{C.~G. Bampis}, \bibinfo{person}{Z. Li},
  \bibinfo{person}{A.~K. Moorthy}, \bibinfo{person}{I. Katsavounidis},
  \bibinfo{person}{A. Aaron}, {and} \bibinfo{person}{A.~C. Bovik}.}
  \bibinfo{year}{2017}\natexlab{}.
\newblock \showarticletitle{Study of Temporal Effects on Subjective Video
  {Quality of Experience}}.
\newblock \bibinfo{journal}{\emph{IEEE Trans. Image Processing}}
  \bibinfo{volume}{26}, \bibinfo{number}{11} (\bibinfo{date}{Nov.}
  \bibinfo{year}{2017}), \bibinfo{pages}{5217--5231}.
\newblock


\bibitem[Bentaleb et~al\mbox{.}(2016)]%
        {bentaleb2016sdndash}
\bibfield{author}{\bibinfo{person}{A. Bentaleb}, \bibinfo{person}{A.~C. Begen},
  {and} \bibinfo{person}{R. Zimmermann}.} \bibinfo{year}{2016}\natexlab{}.
\newblock \showarticletitle{{SDNDASH}: {Improving} {QoE} of {HTTP} adaptive
  streaming using software defined networking}. In
  \bibinfo{booktitle}{\emph{Proc. ACM Int. Conf. Multimedia}}.
  \bibinfo{publisher}{ACM}, \bibinfo{address}{Amsterdam, The Netherlands},
  \bibinfo{pages}{1296--1305}.
\newblock


\bibitem[Bronzino et~al\mbox{.}(2019)]%
        {bronzino2019inferring}
\bibfield{author}{\bibinfo{person}{Francesco Bronzino}, \bibinfo{person}{Paul
  Schmitt}, \bibinfo{person}{Sara Ayoubi}, \bibinfo{person}{Guilherme Martins},
  \bibinfo{person}{Renata Teixeira}, {and} \bibinfo{person}{Nick Feamster}.}
  \bibinfo{year}{2019}\natexlab{}.
\newblock \showarticletitle{Inferring streaming video quality from encrypted
  traffic: Practical models and deployment experience}.
\newblock \bibinfo{journal}{\emph{Proceedings of the ACM on Measurement and
  Analysis of Computing Systems}} \bibinfo{volume}{3}, \bibinfo{number}{3}
  (\bibinfo{year}{2019}), \bibinfo{pages}{1--25}.
\newblock


\bibitem[Chang et~al\mbox{.}(2018)]%
        {chang2018active}
\bibfield{author}{\bibinfo{person}{Haw-Shiuan Chang}, \bibinfo{person}{Chih-Fan
  Hsu}, \bibinfo{person}{Tobias Ho{\ss}feld}, {and} \bibinfo{person}{Kuan-Ta
  Chen}.} \bibinfo{year}{2018}\natexlab{}.
\newblock \showarticletitle{Active learning for crowdsourced QoE modeling}.
\newblock \bibinfo{journal}{\emph{IEEE Transactions on Multimedia}}
  \bibinfo{volume}{20}, \bibinfo{number}{12} (\bibinfo{year}{2018}),
  \bibinfo{pages}{3337--3352}.
\newblock


\bibitem[Chen et~al\mbox{.}(2015)]%
        {chen2015qoe}
\bibfield{author}{\bibinfo{person}{Yanjiao Chen}, \bibinfo{person}{Fan Zhang},
  \bibinfo{person}{Kaishun Wu}, {and} \bibinfo{person}{Qian Zhang}.}
  \bibinfo{year}{2015}\natexlab{}.
\newblock \showarticletitle{Qoe-aware dynamic video rate adaptation}. In
  \bibinfo{booktitle}{\emph{2015 IEEE Global Communications Conference
  (GLOBECOM)}}. IEEE, \bibinfo{pages}{1--6}.
\newblock


\bibitem[Chiariotti et~al\mbox{.}(2016)]%
        {chiariotti2016online}
\bibfield{author}{\bibinfo{person}{F. Chiariotti}, \bibinfo{person}{S.
  D'Aronco}, \bibinfo{person}{L. Toni}, {and} \bibinfo{person}{P. Frossard}.}
  \bibinfo{year}{2016}\natexlab{}.
\newblock \showarticletitle{Online Learning Adaptation Strategy for {DASH}
  Clients}. In \bibinfo{booktitle}{\emph{Proc. ACM Conf. Multimedia Systems}}.
  \bibinfo{publisher}{ACM}, \bibinfo{address}{Klagenfurt, Austria},
  \bibinfo{pages}{1--12}.
\newblock


\bibitem[Davidson(2013)]%
        {convivadamage}
\bibfield{author}{\bibinfo{person}{Neil Davidson}.}
  \bibinfo{year}{2013}\natexlab{}.
\newblock \bibinfo{title}{6 Ways Low Quality Video Production Can Damage Your
  Business}.
\newblock
  \bibinfo{howpublished}{\url{https://www.business2community.com/marketing/6-ways-low-quality-video-production-can-damage-your-business-0424796}}.
\newblock


\bibitem[De~Cock et~al\mbox{.}(2016)]%
        {de2016complexity}
\bibfield{author}{\bibinfo{person}{J. De~Cock}, \bibinfo{person}{Z. Li},
  \bibinfo{person}{M. Manohara}, {and} \bibinfo{person}{A. Aaron}.}
  \bibinfo{year}{2016}\natexlab{}.
\newblock \showarticletitle{Complexity-Based Consistent-Quality Encoding in the
  Cloud}. In \bibinfo{booktitle}{\emph{Proc. IEEE Int. Conf. Image Proc.}}
  \bibinfo{publisher}{IEEE}, \bibinfo{address}{Phoenix, AZ, USA},
  \bibinfo{pages}{1484--1488}.
\newblock


\bibitem[Dobrian et~al\mbox{.}(2011)]%
        {dobrian2011understanding}
\bibfield{author}{\bibinfo{person}{Florin Dobrian}, \bibinfo{person}{Vyas
  Sekar}, \bibinfo{person}{Asad Awan}, \bibinfo{person}{Ion Stoica},
  \bibinfo{person}{Dilip Joseph}, \bibinfo{person}{Aditya Ganjam},
  \bibinfo{person}{Jibin Zhan}, {and} \bibinfo{person}{Hui Zhang}.}
  \bibinfo{year}{2011}\natexlab{}.
\newblock \showarticletitle{Understanding the impact of video quality on user
  engagement}.
\newblock \bibinfo{journal}{\emph{ACM SIGCOMM Computer Communication Review}}
  \bibinfo{volume}{41}, \bibinfo{number}{4} (\bibinfo{year}{2011}),
  \bibinfo{pages}{362--373}.
\newblock


\bibitem[Duanmu et~al\mbox{.}(2019a)]%
        {duanmu2019sqoe4}
\bibfield{author}{\bibinfo{person}{Z. Duanmu}, \bibinfo{person}{D. Chen},
  \bibinfo{person}{Z. Li}, \bibinfo{person}{W. Liu}, \bibinfo{person}{Z. Wang},
  \bibinfo{person}{Y. Wang}, {and} \bibinfo{person}{W. Gao}.}
  \bibinfo{year}{2019}\natexlab{a}.
\newblock \bibinfo{booktitle}{\emph{Waterloo Streaming {Quality-of-Experince}
  Database {IV}}}.
\newblock
\urldef\tempurl%
\url{http://ece.uwaterloo.ca/~zduanmu/waterloosqoe4.}
\showURL{%
\tempurl}


\bibitem[Duanmu et~al\mbox{.}(2019b)]%
        {duanmu2019knowledge}
\bibfield{author}{\bibinfo{person}{Zhengfang Duanmu}, \bibinfo{person}{Wentao
  Liu}, \bibinfo{person}{Diqi Chen}, \bibinfo{person}{Zhuoran Li},
  \bibinfo{person}{Zhou Wang}, \bibinfo{person}{Yizhou Wang}, {and}
  \bibinfo{person}{Wen Gao}.} \bibinfo{year}{2019}\natexlab{b}.
\newblock \showarticletitle{A Knowledge-Driven Quality-of-Experience Model for
  Adaptive Streaming Videos}.
\newblock \bibinfo{journal}{\emph{arXiv preprint arXiv:1911.07944}}
  (\bibinfo{year}{2019}).
\newblock


\bibitem[Duanmu et~al\mbox{.}(2017a)]%
        {duanmu2017qoe}
\bibfield{author}{\bibinfo{person}{Z. Duanmu}, \bibinfo{person}{K. Ma}, {and}
  \bibinfo{person}{Z. Wang}.} \bibinfo{year}{2017}\natexlab{a}.
\newblock \showarticletitle{{Quality-of-Experience} of Adaptive Video
  Streaming: {Exploring} the Space of Adaptations}. In
  \bibinfo{booktitle}{\emph{Proc. ACM Int. Conf. Multimedia}}.
  \bibinfo{publisher}{ACM}, \bibinfo{address}{Mountain View, CA, USA},
  \bibinfo{pages}{1752--1760}.
\newblock


\bibitem[Duanmu et~al\mbox{.}(2018)]%
        {duanmu2018quality}
\bibfield{author}{\bibinfo{person}{Z. Duanmu}, \bibinfo{person}{A. Rehman},
  {and} \bibinfo{person}{Z. Wang}.} \bibinfo{year}{2018}\natexlab{}.
\newblock \showarticletitle{A {Quality-of-Experience} database for adaptive
  video streaming}.
\newblock \bibinfo{journal}{\emph{IEEE Trans. Broadcasting}}
  \bibinfo{volume}{64}, \bibinfo{number}{2} (\bibinfo{date}{Jun.}
  \bibinfo{year}{2018}), \bibinfo{pages}{474--487}.
\newblock


\bibitem[Duanmu et~al\mbox{.}(2017b)]%
        {duanmu2016sqi}
\bibfield{author}{\bibinfo{person}{Z. Duanmu}, \bibinfo{person}{K. Zeng},
  \bibinfo{person}{K. Ma}, \bibinfo{person}{A. Rehman}, {and}
  \bibinfo{person}{Z. Wang}.} \bibinfo{year}{2017}\natexlab{b}.
\newblock \showarticletitle{A {Quality-of-Experience} Index for Streaming
  Video}.
\newblock \bibinfo{journal}{\emph{IEEE Journal of Selected Topics in Signal
  Processing}} \bibinfo{volume}{11}, \bibinfo{number}{1} (\bibinfo{date}{Sep.}
  \bibinfo{year}{2017}), \bibinfo{pages}{154--166}.
\newblock


\bibitem[Ganjam et~al\mbox{.}(2015)]%
        {ganjam2015c3}
\bibfield{author}{\bibinfo{person}{Aditya Ganjam}, \bibinfo{person}{Faisal
  Siddiqui}, \bibinfo{person}{Jibin Zhan}, \bibinfo{person}{Xi Liu},
  \bibinfo{person}{Ion Stoica}, \bibinfo{person}{Junchen Jiang},
  \bibinfo{person}{Vyas Sekar}, {and} \bibinfo{person}{Hui Zhang}.}
  \bibinfo{year}{2015}\natexlab{}.
\newblock \showarticletitle{C3: Internet-scale control plane for video quality
  optimization}. In \bibinfo{booktitle}{\emph{USENIX Symp. Networked Systems
  Design and Implementation}}. \bibinfo{publisher}{{USENIX} Association},
  \bibinfo{address}{Oakland, CA, USA}, \bibinfo{pages}{131--144}.
\newblock


\bibitem[Gao et~al\mbox{.}(pear)]%
        {gao2020personalized}
\bibfield{author}{\bibinfo{person}{Yun Gao}, \bibinfo{person}{Xin Wei}, {and}
  \bibinfo{person}{Liang Zhou}.} \bibinfo{year}{2020, To Appear}\natexlab{}.
\newblock \showarticletitle{Personalized QoE Improvement for Networking Video
  Service}.
\newblock \bibinfo{journal}{\emph{IEEE Journal on Selected Areas in
  Communications}} (\bibinfo{year}{2020, To Appear}).
\newblock


\bibitem[G{\'o}mez et~al\mbox{.}(2013)]%
        {gomez2013towards}
\bibfield{author}{\bibinfo{person}{Gerardo G{\'o}mez}, \bibinfo{person}{Javier
  Lorca}, \bibinfo{person}{Raquel Garc{\'\i}a}, {and} \bibinfo{person}{Quiliano
  P{\'e}rez}.} \bibinfo{year}{2013}\natexlab{}.
\newblock \showarticletitle{Towards a {QoE}-driven resource control in {LTE}
  and {LTE-A} networks}.
\newblock \bibinfo{journal}{\emph{Journal of Computer Networks and
  Communications}}  \bibinfo{volume}{2013} (\bibinfo{year}{2013}).
\newblock


\bibitem[Grafl and Timmerer(2013)]%
        {grafl2013representation}
\bibfield{author}{\bibinfo{person}{M. Grafl} {and} \bibinfo{person}{C.
  Timmerer}.} \bibinfo{year}{2013}\natexlab{}.
\newblock \showarticletitle{Representation Switch Smoothing for Adaptive {HTTP}
  Streaming}. In \bibinfo{booktitle}{\emph{Proc. IEEE Int. Workshop Perceptual
  Qualtiy of Systems}}. \bibinfo{publisher}{ISCA/DEGA},
  \bibinfo{address}{Vienna, Austria}, \bibinfo{pages}{178--183}.
\newblock


\bibitem[Gr{\"u}ner et~al\mbox{.}(2020)]%
        {gruner2020reconstructing}
\bibfield{author}{\bibinfo{person}{Maximilian Gr{\"u}ner},
  \bibinfo{person}{Melissa Licciardello}, {and} \bibinfo{person}{Ankit
  Singla}.} \bibinfo{year}{2020}\natexlab{}.
\newblock \showarticletitle{Reconstructing proprietary video streaming
  algorithms}. In \bibinfo{booktitle}{\emph{2020 USENIX Annual Technical
  Conference (USENIX ATC 20)}}.
\newblock


\bibitem[Hootsuite(2021)]%
        {youtubestats}
\bibfield{author}{\bibinfo{person}{Hootsuite}.}
  \bibinfo{year}{2021}\natexlab{}.
\newblock \bibinfo{title}{25 YouTube Statistics that May Surprise You: 2021
  Edition}.
\newblock
  \bibinfo{howpublished}{\url{https://blog.hootsuite.com/youtube-stats-marketers/}}.
\newblock


\bibitem[Ho{\ss}feld et~al\mbox{.}(2013)]%
        {tobias2013youtube}
\bibfield{author}{\bibinfo{person}{T. Ho{\ss}feld}, \bibinfo{person}{R.
  Schatz}, \bibinfo{person}{E. Biersack}, {and} \bibinfo{person}{L.
  Plissonneau}.} \bibinfo{year}{2013}\natexlab{}.
\newblock \showarticletitle{Internet Video Delivery in {YouTube}: {From}
  Traffic Measurements to {Q}uality of {E}xperience}.
\newblock In \bibinfo{booktitle}{\emph{Data Traffic Monitoring and Analysis}}.
  \bibinfo{publisher}{Springer}, \bibinfo{address}{Berlin, Heidelberg},
  \bibinfo{pages}{264--301}.
\newblock


\bibitem[Ho{\ss}feld et~al\mbox{.}(2011)]%
        {tobias2011youtube}
\bibfield{author}{\bibinfo{person}{T. Ho{\ss}feld}, \bibinfo{person}{M.
  Seufert}, \bibinfo{person}{M. Hirth}, \bibinfo{person}{T. Zinner},
  \bibinfo{person}{P. Tran-Gia}, {and} \bibinfo{person}{R. Schatz}.}
  \bibinfo{year}{2011}\natexlab{}.
\newblock \showarticletitle{Quantification of {YouTube} {QoE} via
  Crowdsourcing}. In \bibinfo{booktitle}{\emph{Proc. IEEE Int. Sym.
  Multimedia}}. \bibinfo{publisher}{IEEE}, \bibinfo{address}{Dana Point, CA,
  USA}, \bibinfo{pages}{494--499}.
\newblock


\bibitem[{Huo} et~al\mbox{.}(pear)]%
        {huo2020meta}
\bibfield{author}{\bibinfo{person}{L. {Huo}}, \bibinfo{person}{Z. {Wang}},
  \bibinfo{person}{M. {Xu}}, \bibinfo{person}{Y. {Li}}, \bibinfo{person}{Z.
  {Ding}}, {and} \bibinfo{person}{H. {Wang}}.} \bibinfo{year}{2020, To
  Appear}\natexlab{}.
\newblock \showarticletitle{A Meta-learning Framework for Learning Multi-User
  Preferences in {QoE} Optimization of {DASH}}.
\newblock \bibinfo{journal}{\emph{IEEE Transactions on Circuits and Systems for
  Video Technology}} (\bibinfo{year}{2020, To Appear}), \bibinfo{pages}{1--1}.
\newblock


\bibitem[{ITU-R BT.500-12}(1993)]%
        {bt500subjective}
\bibfield{author}{\bibinfo{person}{{ITU-R BT.500-12}}.}
  \bibinfo{year}{1993}\natexlab{}.
\newblock \bibinfo{title}{Recommendation: {Methodology} for the Subjective
  Assessment of the Quality of Television Pictures}.
\newblock
\newblock


\bibitem[Jiang et~al\mbox{.}(2016)]%
        {jiang2016cfa}
\bibfield{author}{\bibinfo{person}{Junchen Jiang}, \bibinfo{person}{Vyas
  Sekar}, \bibinfo{person}{Henry Milner}, \bibinfo{person}{Davis Shepherd},
  \bibinfo{person}{Ion Stoica}, {and} \bibinfo{person}{Hui Zhang}.}
  \bibinfo{year}{2016}\natexlab{}.
\newblock \showarticletitle{{CFA}: {A} Practical Prediction System for Video
  {QoE} Optimization}. In \bibinfo{booktitle}{\emph{USENIX Symp. Networked
  Systems Design and Implementation}}. \bibinfo{publisher}{{USENIX}
  Association}, \bibinfo{address}{Santa Clara, CA, USA},
  \bibinfo{pages}{137--150}.
\newblock


\bibitem[Jiang et~al\mbox{.}(2017)]%
        {jiang2017pytheas}
\bibfield{author}{\bibinfo{person}{Junchen Jiang}, \bibinfo{person}{Shijie
  Sun}, \bibinfo{person}{Vyas Sekar}, {and} \bibinfo{person}{Hui Zhang}.}
  \bibinfo{year}{2017}\natexlab{}.
\newblock \showarticletitle{Pytheas: Enabling Data-Driven Quality of Experience
  Optimization Using Group-Based Exploration-Exploitation}. In
  \bibinfo{booktitle}{\emph{USENIX Symp. Networked Systems Design and
  Implementation}}. \bibinfo{publisher}{{USENIX} Association},
  \bibinfo{address}{Boston, MA, USA}, \bibinfo{pages}{393--406}.
\newblock


\bibitem[Jiang et~al\mbox{.}(2012)]%
        {jiang2012qoe}
\bibfield{author}{\bibinfo{person}{Tigang Jiang}, \bibinfo{person}{Honggang
  Wang}, {and} \bibinfo{person}{Athanasios~V Vasilakos}.}
  \bibinfo{year}{2012}\natexlab{}.
\newblock \showarticletitle{{QoE}-driven channel allocation schemes for
  multimedia transmission of priority-based secondary users over cognitive
  radio networks}.
\newblock \bibinfo{journal}{\emph{IEEE Journal on Selected Areas in
  Communications}} \bibinfo{volume}{30}, \bibinfo{number}{7}
  (\bibinfo{year}{2012}), \bibinfo{pages}{1215--1224}.
\newblock


\bibitem[Li et~al\mbox{.}(2016)]%
        {li2016VMAF}
\bibfield{author}{\bibinfo{person}{Z. Li}, \bibinfo{person}{A. Aaron},
  \bibinfo{person}{L. Katsavounidis}, \bibinfo{person}{A. Moorthy}, {and}
  \bibinfo{person}{M. Manohara}.} \bibinfo{year}{2016}\natexlab{}.
\newblock \bibinfo{booktitle}{\emph{Toward A Practical Perceptual Video Quality
  Metric}}.
\newblock
\urldef\tempurl%
\url{http://techblog.netflix.com/2016/06/toward-practical-perceptual-video.html.}
\showURL{%
Retrieved July 20, 2018 from \tempurl}


\bibitem[Liu et~al\mbox{.}(2012)]%
        {liu2012case}
\bibfield{author}{\bibinfo{person}{X. Liu}, \bibinfo{person}{F. Dobrian},
  \bibinfo{person}{H. Milner}, \bibinfo{person}{J. Jiang}, \bibinfo{person}{V.
  Sekar}, \bibinfo{person}{I. Stoica}, {and} \bibinfo{person}{H. Zhang}.}
  \bibinfo{year}{2012}\natexlab{}.
\newblock \showarticletitle{A case for a coordinated internet video control
  plane}.
\newblock \bibinfo{journal}{\emph{ACM SIGCOMM Computer Communication Review}}
  \bibinfo{volume}{42}, \bibinfo{number}{4} (\bibinfo{date}{Sep.}
  \bibinfo{year}{2012}), \bibinfo{pages}{359--370}.
\newblock


\bibitem[Loomes and Sugden(1982)]%
        {loomes1982regret}
\bibfield{author}{\bibinfo{person}{Graham Loomes} {and} \bibinfo{person}{Robert
  Sugden}.} \bibinfo{year}{1982}\natexlab{}.
\newblock \showarticletitle{Regret theory: An alternative theory of rational
  choice under uncertainty}.
\newblock \bibinfo{journal}{\emph{The economic journal}} \bibinfo{volume}{92},
  \bibinfo{number}{368} (\bibinfo{year}{1982}), \bibinfo{pages}{805--824}.
\newblock


\bibitem[Ma et~al\mbox{.}(2020)]%
        {ma2019gmad}
\bibfield{author}{\bibinfo{person}{K. Ma}, \bibinfo{person}{Z. Duanmu},
  \bibinfo{person}{Z. Wang}, \bibinfo{person}{Q. Wu}, \bibinfo{person}{W. Liu},
  \bibinfo{person}{H. Yong}, \bibinfo{person}{H. Li}, {and} \bibinfo{person}{L.
  Zhang}.} \bibinfo{year}{2020}\natexlab{}.
\newblock \showarticletitle{Group Maximum Differentiation Competition: {Model}
  Comparison with Few Samples}.
\newblock \bibinfo{journal}{\emph{IEEE Trans. Pattern Analysis and Machine
  Intelligence}} \bibinfo{volume}{42}, \bibinfo{number}{4}
  (\bibinfo{year}{2020}), \bibinfo{pages}{851--864}.
\newblock


\bibitem[Mao et~al\mbox{.}(2017)]%
        {mao2017neural}
\bibfield{author}{\bibinfo{person}{H. Mao}, \bibinfo{person}{R. Netravali},
  {and} \bibinfo{person}{M. Alizadeh}.} \bibinfo{year}{2017}\natexlab{}.
\newblock \showarticletitle{Neural Adaptive Video Streaming with {Pensieve}}.
  In \bibinfo{booktitle}{\emph{Proc. ACM SIGCOMM}}. \bibinfo{publisher}{ACM},
  \bibinfo{address}{Los Angeles, CA, USA}, \bibinfo{pages}{197--210}.
\newblock


\bibitem[Menkovski et~al\mbox{.}(2010)]%
        {menkovski2010online}
\bibfield{author}{\bibinfo{person}{Vlado Menkovski}, \bibinfo{person}{Georgios
  Exarchakos}, {and} \bibinfo{person}{Antonio Liotta}.}
  \bibinfo{year}{2010}\natexlab{}.
\newblock \showarticletitle{Online learning for quality of experience
  management}. In \bibinfo{booktitle}{\emph{Annual Machine Learning Conference
  of Belgium and The Netherlands}}. \bibinfo{pages}{6}.
\newblock


\bibitem[Mok et~al\mbox{.}(2012)]%
        {mok2012qdash}
\bibfield{author}{\bibinfo{person}{R.~K. Mok}, \bibinfo{person}{X. Luo},
  \bibinfo{person}{E.~W. Chan}, {and} \bibinfo{person}{R.~K. Chang}.}
  \bibinfo{year}{2012}\natexlab{}.
\newblock \showarticletitle{{QDASH}: A {QoE}-Aware {DASH} System}. In
  \bibinfo{booktitle}{\emph{Proc. ACM Conf. Multimedia Systems}}.
  \bibinfo{publisher}{ACM}, \bibinfo{address}{Chapel Hill, NC, USA},
  \bibinfo{pages}{11--22}.
\newblock


\bibitem[Moorthy et~al\mbox{.}(2012)]%
        {moorthy2012video}
\bibfield{author}{\bibinfo{person}{A.~K. Moorthy}, \bibinfo{person}{L.~K.
  Choi}, \bibinfo{person}{A.~C. Bovik}, {and} \bibinfo{person}{G. De~Veciana}.}
  \bibinfo{year}{2012}\natexlab{}.
\newblock \showarticletitle{Video Quality Assessment on Mobile Devices:
  {Subjective}, Behavioral and Objective Studies}.
\newblock \bibinfo{journal}{\emph{IEEE Journal of Selected Topics in Signal
  Processing}} \bibinfo{volume}{6}, \bibinfo{number}{6} (\bibinfo{date}{Oct.}
  \bibinfo{year}{2012}), \bibinfo{pages}{652--671}.
\newblock


\bibitem[Mozilla(2022)]%
        {mediasourceAPI}
\bibfield{author}{\bibinfo{person}{Mozilla}.} \bibinfo{year}{2022}\natexlab{}.
\newblock \bibinfo{title}{Media Source Extensions (MSE)}.
\newblock
  \bibinfo{howpublished}{\url{https://developer.mozilla.org/en-US/docs/Web/API/Media_Source_Extensions_API}}.
\newblock


\bibitem[MPEG(2015)]%
        {fragmentmp4}
\bibfield{author}{\bibinfo{person}{MPEG}.} \bibinfo{year}{2015}\natexlab{}.
\newblock \bibinfo{title}{Text of ISO/IEC 14496-12 5th edition}.
\newblock
  \bibinfo{howpublished}{\url{https://mpeg.chiariglione.org/standards/mpeg-4/iso-base-media-file-format/text-isoiec-14496-12-5th-edition}}.
\newblock


\bibitem[Nathan et~al\mbox{.}(2019)]%
        {nathan2019end}
\bibfield{author}{\bibinfo{person}{Vikram Nathan},
  \bibinfo{person}{Vibhaalakshmi Sivaraman}, \bibinfo{person}{Ravichandra
  Addanki}, \bibinfo{person}{Mehrdad Khani}, \bibinfo{person}{Prateesh Goyal},
  {and} \bibinfo{person}{Mohammad Alizadeh}.} \bibinfo{year}{2019}\natexlab{}.
\newblock \showarticletitle{End-to-end transport for video QoE fairness}.
\newblock In \bibinfo{booktitle}{\emph{Proceedings of the ACM Special Interest
  Group on Data Communication}}. \bibinfo{pages}{408--423}.
\newblock


\bibitem[Ni et~al\mbox{.}(2011)]%
        {ni2011flicker}
\bibfield{author}{\bibinfo{person}{P. Ni}, \bibinfo{person}{R. Eg},
  \bibinfo{person}{A. Eichhorn}, \bibinfo{person}{C. Griwodz}, {and}
  \bibinfo{person}{P. Halvorsen}.} \bibinfo{year}{2011}\natexlab{}.
\newblock \showarticletitle{Flicker effects in adaptive video streaming to
  handheld devices}. In \bibinfo{booktitle}{\emph{Proc. ACM Int. Conf.
  Multimedia}}. \bibinfo{publisher}{ACM}, \bibinfo{address}{Scottsdale, AZ,
  USA}, \bibinfo{pages}{463--472}.
\newblock


\bibitem[P.1203(2017)]%
        {itu2017pnats}
\bibfield{author}{\bibinfo{person}{ITU-T P.1203}.}
  \bibinfo{year}{2017}\natexlab{}.
\newblock \bibinfo{booktitle}{\emph{Parametric bitstream-based quality
  assessment of progressive download and adaptive audiovisual streaming
  services over reliable transport}}.
\newblock
\urldef\tempurl%
\url{https://www.itu.int/rec/dologin_pub.asp?lang=e&id=T-REC-P.1203-201710-I!!PDF-E&type=items.}
\showURL{%
\tempurl}


\bibitem[Pastrana-Vidal et~al\mbox{.}(2004)]%
        {pastrana2004sporadic}
\bibfield{author}{\bibinfo{person}{R. Pastrana-Vidal}, \bibinfo{person}{J.~C.
  Gicquel}, \bibinfo{person}{C. Colomes}, {and} \bibinfo{person}{H. Cherifi}.}
  \bibinfo{year}{2004}\natexlab{}.
\newblock \showarticletitle{Sporadic Frame Dropping Impact on Quality
  Perception}. In \bibinfo{booktitle}{\emph{Human Vision and Electronic Imaging
  IX}}. \bibinfo{publisher}{SPIE}, \bibinfo{address}{San Jose, CA, USA},
  \bibinfo{pages}{182--194}.
\newblock


\bibitem[Prolifc(2022)]%
        {prolificlink}
\bibfield{author}{\bibinfo{person}{Prolifc}.} \bibinfo{year}{2022}\natexlab{}.
\newblock \bibinfo{howpublished}{\url{https://prolific.co/}}.
\newblock


\bibitem[Qi and Dai(2006)]%
        {qi2006effect}
\bibfield{author}{\bibinfo{person}{Y. Qi} {and} \bibinfo{person}{M. Dai}.}
  \bibinfo{year}{2006}\natexlab{}.
\newblock \showarticletitle{The Effect of Frame Freezing and Frame Skipping on
  Video Quality}. In \bibinfo{booktitle}{\emph{Proc. IEEE Int. Conf.
  Intelligent Information Hiding and Multimedia Signal Processing}}.
  \bibinfo{publisher}{IEEE}, \bibinfo{address}{Pasadena, CA, USA},
  \bibinfo{pages}{423--426}.
\newblock


\bibitem[Rehman and Wang(2013)]%
        {rehman2013perceptual}
\bibfield{author}{\bibinfo{person}{A. Rehman} {and} \bibinfo{person}{Z. Wang}.}
  \bibinfo{year}{2013}\natexlab{}.
\newblock \showarticletitle{Perceptual Experience of Time-Varying Video
  Quality}. In \bibinfo{booktitle}{\emph{Proc. IEEE Int. Conf. Quality of
  Multimedia Experience}}. \bibinfo{publisher}{IEEE},
  \bibinfo{address}{Klagenfurt am Worthersee}, \bibinfo{pages}{218--223}.
\newblock


\bibitem[Rehman et~al\mbox{.}(2015)]%
        {rehman2015ssimplus}
\bibfield{author}{\bibinfo{person}{A. Rehman}, \bibinfo{person}{K. Zeng}, {and}
  \bibinfo{person}{Z. Wang}.} \bibinfo{year}{2015}\natexlab{}.
\newblock \showarticletitle{Display Device-Adapted Video
  {Quality-of-Experience} Assessment}. In \bibinfo{booktitle}{\emph{Proc.
  SPIE}}. \bibinfo{publisher}{SPIE}, \bibinfo{address}{San Francisco, CA, USA},
  \bibinfo{pages}{939406.1--939406.11}.
\newblock


\bibitem[Settles(2009)]%
        {settles2009active}
\bibfield{author}{\bibinfo{person}{Burr Settles}.}
  \bibinfo{year}{2009}\natexlab{}.
\newblock \showarticletitle{Active learning literature survey}.
\newblock  (\bibinfo{year}{2009}).
\newblock


\bibitem[Spiteri et~al\mbox{.}(2016)]%
        {spiteri2016bola}
\bibfield{author}{\bibinfo{person}{K. Spiteri}, \bibinfo{person}{R. Urgaonkar},
  {and} \bibinfo{person}{R.~K. Sitaraman}.} \bibinfo{year}{2016}\natexlab{}.
\newblock \showarticletitle{{BOLA}: {Near}-Optimal Bitrate Adaptation for
  Online Videos}. In \bibinfo{booktitle}{\emph{Proc. IEEE Int. Conf. Computer
  Communications}}. \bibinfo{publisher}{IEEE}, \bibinfo{address}{San Francisco,
  CA, USA}, \bibinfo{pages}{1--9}.
\newblock


\bibitem[Staelens et~al\mbox{.}(2010)]%
        {staelens2010assessing}
\bibfield{author}{\bibinfo{person}{N. Staelens}, \bibinfo{person}{S. Moens},
  \bibinfo{person}{W.~Van den Broeck}, \bibinfo{person}{I. Marien},
  \bibinfo{person}{B. Vermeulen}, \bibinfo{person}{P. Lambert},
  \bibinfo{person}{R.~Van de Walle}, {and} \bibinfo{person}{P. Demeester}.}
  \bibinfo{year}{2010}\natexlab{}.
\newblock \showarticletitle{Assessing {Quality of Experience} of {IPTV} and
  Video on Demand Services in Real-Life Environments}.
\newblock \bibinfo{journal}{\emph{IEEE Trans. Broadcasting}}
  \bibinfo{volume}{56}, \bibinfo{number}{4} (\bibinfo{date}{Dec.}
  \bibinfo{year}{2010}), \bibinfo{pages}{458--466}.
\newblock


\bibitem[Toni et~al\mbox{.}(2015)]%
        {toni2015optimal}
\bibfield{author}{\bibinfo{person}{L. Toni}, \bibinfo{person}{R.
  Aparicio-Pardo}, \bibinfo{person}{K. Pires}, \bibinfo{person}{W. Simon},
  \bibinfo{person}{A. Blanc}, {and} \bibinfo{person}{P. Frossard}.}
  \bibinfo{year}{2015}\natexlab{}.
\newblock \showarticletitle{Optimal selection of adaptive streaming
  representations}.
\newblock \bibinfo{journal}{\emph{ACM Trans. Multimedia Computing,
  Communications, and Applications}} \bibinfo{volume}{11}, \bibinfo{number}{2s}
  (\bibinfo{date}{Feb.} \bibinfo{year}{2015}), \bibinfo{pages}{1--43}.
\newblock


\bibitem[Wang et~al\mbox{.}(2016)]%
        {wang2016data}
\bibfield{author}{\bibinfo{person}{Ying Wang}, \bibinfo{person}{Peilong Li},
  \bibinfo{person}{Lei Jiao}, \bibinfo{person}{Zhou Su}, \bibinfo{person}{Nan
  Cheng}, \bibinfo{person}{Xuemin~Sherman Shen}, {and} \bibinfo{person}{Ping
  Zhang}.} \bibinfo{year}{2016}\natexlab{}.
\newblock \showarticletitle{A data-driven architecture for personalized {QoE}
  management in {5G} wireless networks}.
\newblock \bibinfo{journal}{\emph{IEEE Wireless Communications}}
  \bibinfo{volume}{24}, \bibinfo{number}{1} (\bibinfo{year}{2016}),
  \bibinfo{pages}{102--110}.
\newblock


\bibitem[Yan et~al\mbox{.}(2020)]%
        {yan2020learning}
\bibfield{author}{\bibinfo{person}{Francis~Y Yan}, \bibinfo{person}{Hudson
  Ayers}, \bibinfo{person}{Chenzhi Zhu}, \bibinfo{person}{Sadjad Fouladi},
  \bibinfo{person}{James Hong}, \bibinfo{person}{Keyi Zhang},
  \bibinfo{person}{Philip Levis}, {and} \bibinfo{person}{Keith Winstein}.}
  \bibinfo{year}{2020}\natexlab{}.
\newblock \showarticletitle{Learning in situ: a randomized experiment in video
  streaming}. In \bibinfo{booktitle}{\emph{17th $\{$USENIX$\}$ Symposium on
  Networked Systems Design and Implementation ($\{$NSDI$\}$ 20)}}.
  \bibinfo{pages}{495--511}.
\newblock


\bibitem[Yin et~al\mbox{.}(2015)]%
        {yin2015control}
\bibfield{author}{\bibinfo{person}{X. Yin}, \bibinfo{person}{A. Jindal},
  \bibinfo{person}{V. Sekar}, {and} \bibinfo{person}{B. Sinopoli}.}
  \bibinfo{year}{2015}\natexlab{}.
\newblock \showarticletitle{A Control-Theoretic Approach for Dynamic Adaptive
  Video Streaming over {HTTP}}.
\newblock \bibinfo{journal}{\emph{ACM SIGCOMM Computer Communication Review}}
  \bibinfo{volume}{45}, \bibinfo{number}{4} (\bibinfo{date}{Apr.}
  \bibinfo{year}{2015}), \bibinfo{pages}{325--338}.
\newblock


\bibitem[Young(2014)]%
        {young2014improving}
\bibfield{author}{\bibinfo{person}{Scott~WH Young}.}
  \bibinfo{year}{2014}\natexlab{}.
\newblock \showarticletitle{Improving library user experience with A/B testing:
  Principles and process}.
\newblock \bibinfo{journal}{\emph{Weave: Journal of Library User Experience}}
  \bibinfo{volume}{1}, \bibinfo{number}{1} (\bibinfo{year}{2014}).
\newblock


\bibitem[YouTube(2022)]%
        {iframeplayer}
\bibfield{author}{\bibinfo{person}{YouTube}.} \bibinfo{year}{2022}\natexlab{}.
\newblock \bibinfo{title}{IFrame Player API}.
\newblock
  \bibinfo{howpublished}{\url{https://developers.google.com/youtube/iframe_api_reference}}.
\newblock


\bibitem[Zhang et~al\mbox{.}(2021)]%
        {zhang2021sensei}
\bibfield{author}{\bibinfo{person}{Xu Zhang}, \bibinfo{person}{Yiyang Ou},
  \bibinfo{person}{Siddhartha Sen}, {and} \bibinfo{person}{Junchen Jiang}.}
  \bibinfo{year}{2021}\natexlab{}.
\newblock \showarticletitle{SENSEI: Aligning Video Streaming Quality with
  Dynamic User Sensitivity}. In \bibinfo{booktitle}{\emph{18th USENIX Symposium
  on Networked Systems Design and Implementation (NSDI 21)}}.
  \bibinfo{pages}{303--320}.
\newblock


\bibitem[Zhao et~al\mbox{.}(2019)]%
        {zhao2019qoe}
\bibfield{author}{\bibinfo{person}{Jialin Zhao}, \bibinfo{person}{Xin Wei},
  \bibinfo{person}{Yun Gao}, \bibinfo{person}{Wenqin Zhuang},
  \bibinfo{person}{Faming Yin}, {and} \bibinfo{person}{Qingbo Du}.}
  \bibinfo{year}{2019}\natexlab{}.
\newblock \showarticletitle{QoE prediction model with personalized parameters
  in IPTV domain}. In \bibinfo{booktitle}{\emph{2019 IEEE International
  Conference on Consumer Electronics-Taiwan (ICCE-TW)}}. IEEE,
  \bibinfo{pages}{1--2}.
\newblock


\end{thebibliography}

\appendix

\section{Client-side browser extension}
\label{sec:impl:plugin}


In this paper, we only test \name when the users watch YouTube videos. 
We strictly follow the settings in \S\ref{sec:design}: we do not change the user interface (UI) of the video streaming service, the users can freely choose their video content to watch for providing us with more natural feedback, and we do not collect the information about video content.

\mypara{Video quality monitor}
We monitor video quality and video session by appending JavaScript codes to the YouTube webpage.
The appended codes have the access to the objects in the webpages.
Thus, we can access the YouTube player object for extracting video download resolution by {\em getPlaybackQuality} API provided by YouTube Iframe player~\cite{iframeplayer}.
But, YouTube does not offer video buffer length information.
To solve this problem, we cast the object of YouTube player to a HTML5 player where we can call {\em SourceBuffer}, one of media source APIs~\cite{mediasourceAPI}, to read the ranges of the video covered by the buffer.

The length of the video session is calculated using the time spend from video session start to when the video ends or the session is closed (whichever comes the first).
Since the users are not in a controlled environment, we do not know the user status when the users watch videos.
To make sure they do watch the videos, we also monitor the browser status, \ie whether the browser is on the top of desktop, and whether the video player tab is show up on the top of the browser.
We only use the videos where the users put the browser on the top of the desktop and the video player tab on the top of the browser to train per-user QoE models, because other videos might not reflect the user preferences to the video quality.

This monitor reports to \name every 0.5 seconds in our implementation.
Each report is <0.1KB, and does not introduce too much traffic to \name's server.

\mypara{Dynamic rate limiting}
We limit the video downloading rate by delaying the HTTP requests whose destination is the video player.
When the HTTP requests arrive at the user client, the bandwidth controller holds the request until it meets the requirement of bandwidth throttling by \emph{onHeadersReceived} API provided by Google Chrome browser.
The time length that the controller needs to hold for a target bandwidth is
\begin{align}
	hold\_time &= \frac{size}{target\_bw} - \frac{size}{real\_bw}\notag,
\end{align}
where $size$ is the data size of the HTTP request, $target\_bw$ is the bandwidth we want to achieve, and the $real\_bw$ is the current real bandwidth of the user.
Although we cannot obtain the ground truth of the real bandwidth, we estimate it as the real bandwidth of the last HTTP request.

Another question is how to determine the hold time of the HTTP requests.
The hold time should make sure the target bandwidth the same as the video bitrate.
To obtain the video bitrate, we can parse the headers of the videos that are mostly using fragmented MP4 format~\cite{fragmentmp4}.
In fragmented MP4, the SIDX field in the header has the byte range of each video chunk in the video streaming, and we can use it to calculate the video bitrate.
We can download the video header for every resolution video source to calculate the bitrate of each version of the video streaming.


\section{Player simulator}
\label{sec:simulator}
Similar to building the player behavior profile in~\S\ref{subsec:dynamicratelimiting}, we use a trace-driven method to simulate a user's player behavior in the video sessions.
Suppose the player is currently at a status with birate $b$, buffer length $l$, the average bandwidth that the player has experienced in the last 20 seconds of the video session $bw_{past}$, and the current bandwidth $bw_{now}$.
We find the historical video sessions that have experienced similar status, \ie in the video session, we can find a player state whose the differences between its bitrate, buffer length, past and current bandwidth, and $b, l, bw_{past}, bw_{now}$  are smaller than $200$Kbps, $5$ seconds and $200$Kbps, $200$Kbps respectively.
We randomly select one of those video sessions and use its player status as the output of the player simulator.


\end{document}